\pdfoutput=1
\documentclass[reprint, superscriptaddress, amsmath,prl,amssymb, aps, floatfix]{revtex4-2}
\makeatletter
\newsavebox{\@brx}
\newcommand{\llangle}[1][]{\savebox{\@brx}{\(\m@th{#1\langle}\)}%
  \mathopen{\copy\@brx\kern-0.5\wd\@brx\usebox{\@brx}}}
\newcommand{\rrangle}[1][]{\savebox{\@brx}{\(\m@th{#1\rangle}\)}%
  \mathclose{\copy\@brx\kern-0.5\wd\@brx\usebox{\@brx}}}
\usepackage{lipsum}
\usepackage{graphicx}
\usepackage{dcolumn}
\usepackage{bm}
\usepackage{xcolor}
\usepackage{xspace}
\usepackage{comment}
\usepackage{braket}
\usepackage{bbm}
\usepackage[colorlinks,allcolors=blue]{hyperref}
\usepackage[makeroom]{cancel}
\usepackage{float}
\usepackage{siunitx}
\usepackage{color}
\definecolor{dgreen}{rgb}{0,0.7,0}
\def\bluew#1{{\color{red} #1}}

\def\bluew#1{{\color{black} #1}}
\newcommand{\beq}{\begin{equation}}
\newcommand{\eeq}{\end{equation}}
\newcommand{\bea}{\begin{eqnarray}}
\newcommand{\eea}{\end{eqnarray}}

\usepackage{subfigure}
\usepackage{booktabs} 
\usepackage{natbib}
\usepackage[normalem]{ulem}

\begin{document}

\title{Transition in Splitting Probabilities of Quantum Walks}
\author{Prashant Singh} 
\email{prashantsinghramitay@gmail.com}
\author{David A. Kessler}
\address{Department of Physics, Bar-Ilan University, Ramat Gan 52900, Israel}
\author{Eli Barkai} 
\address{Department of Physics, Bar-Ilan University, Ramat Gan 52900, Israel}
\address{Institute of Nanotechnology and Advanced Materials, Bar-Ilan University, Ramat Gan 52900, Israel}

\vspace{10pt}

\begin{abstract}
We investigate the splitting probability of a monitored continuous-time quantum walk with two targets and show that, in stark contrast to a classical random walk, it exhibits a nonanalytic, phase-transition-like behavior controlled by the sampling time at the targets. For large systems and sampling times smaller than a critical value $\tau_c = 2\pi/\Delta E$, where $\Delta E$ is the energy bandwidth, the splitting probability is universal and equal to $1/2$, independent of the initial condition and the sampling time. Above the critical sampling time, a nonuniversal regime emerges in which the splitting probability deviates from $1/2$ and develops a fluctuating pattern of pronounced peaks and dips dependent on both the sampling time and the initial condition. These results follow from a nontrivial mapping of the splitting problem onto a pair of single-target detection problems enabled by the superposition principle.

\end{abstract}

\maketitle

\noindent
\textit{Introduction:} Imagine a gambler entering a fair game with a finite initial fortune and aiming to win a predetermined amount, which defines success. At each bet, they gain or lose a part of their current fortune with equal probability. Chance governs every move, yet fate is unavoidable: the gambler either wins or goes bankrupt \cite{shoesmith1986huygens}. What is the probability that they end up winning rather than losing? This question lies at the heart of the \textit{gambler's ruin problem}, where one seeks the splitting probability of achieving one outcome before the other. 

The formulation admits a natural interpretation in terms of a one-dimensional unbiased random walk moving along a finite lattice chain with $N$ number of sites and initial position $x_0$ (with $x_0 \in \{1,2, \ldots N \}$) \cite{Fellerbook, Rednerbook, STSM-2010}. The left and right endpoints of the chain, $x_L = 1$ and $x_R = N$, act as absorbing boundaries, so that the walker is absorbed once it reaches an endpoint and the process stops. 
Eventually, two outcomes are possible: the walker is absorbed at either the left or the right boundary, analogous to winning or losing in the gambler's ruin problem.

The splitting probabilities, $P_L^{\rm (cl)}(x_0)$ and $P_R^{\rm (cl)}(x_0)$, to the left and the right boundaries are \cite{Rednerbook}
\begin{align}
P_L^{\rm (cl)}(x_0) &= \frac{ \left( N - x_0 \right)}{\left( N - 1 \right)}, ~~
P_R^{\rm (cl)}(x_0) = 1 -P_L^{\rm (cl)}(x_0),
\label{lett-unitary-eqn-classical}
\end{align}
where ``cl" denotes the classical nature of the process. The probabilities depend linearly on the initial position. Moreover, the classical walker obeys the proximity effect: the closer the walker starts to a boundary, the more likely it is to be absorbed there. The splitting-probability framework has found wide applications in biophysical systems \cite{ST2008, A1_2011, A1_2015}, stochastic thermodynamics \cite{Neri2015}, polymer physics \cite{A1_2009}, stochastic transport processes \cite{ST-2022, Vezzani2024, nt-ctt-1, disorder1} and restarted processes \cite{J1, nt-ctt-2, nt-ctt-3}.


A plurality of possible outcomes can also have far-reaching implications in quantum settings, motivated by
the expanding scope of quantum applications. For example, in quantum communication, spin-chain-based quantum wires are explicitly used to transmit information from one end of the wire to the other \cite{Qcom2, Qcom1, Bottareli2023, SRPS-5}. Their performance depends on how well the signal reaches the intended end rather than being reflected back. Similarly, quantum walk-based algorithms often involve search processes featuring many targets \cite{Childs2009, Nayak2011,Qcom3, Guan2021HHLQuantumWalk,Qcom4}, and in quantum circuits, the gate operations need to be completed before qubits are irreversibly lost due to noise \cite{Lu2008}.


\bluew{At a fundamental level, such considerations naturally lead to quantum walks with two absorbing boundaries, which provide a minimal setting for analyzing competing outcomes. Previous works, for instance, considered the discrete-time Hadamard walks \cite{Vishwanath2001, SP1, Bach2009, SP5, Ammara2025AbsorbingBoundaries}, a point we return to later.} Physically, absorbing boundaries represent measurement events \cite{Kempe2003,Kempe2-2003, QW1, PSen1, QW25}, and in quantum mechanics, measurement is not a passive readout but rather a dynamical intervention that nontrivially reshapes the system's evolution \cite{Cohenbook}. \bluew{Quantum splitting probabilities therefore are not intrinsic to the system dynamics alone, but are also governed by these dynamical measurement interventions.} Indeed, the sampling time, as defined later, serves as a tunable control parameter in state-of-the-art experiments based on quantum computers \cite{ExptQC-1, ExptQC-2, ExptQC-3}, optical waveguides \cite{ExptWG-1, ExptWG-3, Tang2018, ExptWG-2, Liuexpt1} and trapped ions \cite{Ryan2025}, which so far study the single-target detection problem \cite{QW21,  QW7, QW8, QW6, QW12,QW13, QW16, Das_20221, QW26,Gambassi2025,Perfetto1,Gambassi2026}.

How do quantum measurements shape outcomes among multiple competing targets? Which physical principles govern the competition? What nonclassical features emerge due to quantum fluctuations? The purpose of this Letter is to build an exact theory for the splitting probability of monitored \textit{continuous-time} quantum walks. For a broad class of tight-binding Hamiltonians, we expose a quantum-superposition induced mapping that connects the dual-target splitting problem to a pair of single-target detection problems. This mapping reveals (i) nonanalytic discontinuities of the splitting probability for special sampling times, (ii) a breakdown of the proximity effect, and (iii) the emergence of a phase-transition-like behavior at a critical sampling time, separating two qualitatively distinct splitting regimes. All these features stand in sharp contrast to the classical case in Eq.~\eqref{lett-unitary-eqn-classical}.

\begin{figure}[t]
	\centering
	\includegraphics[scale=0.22]{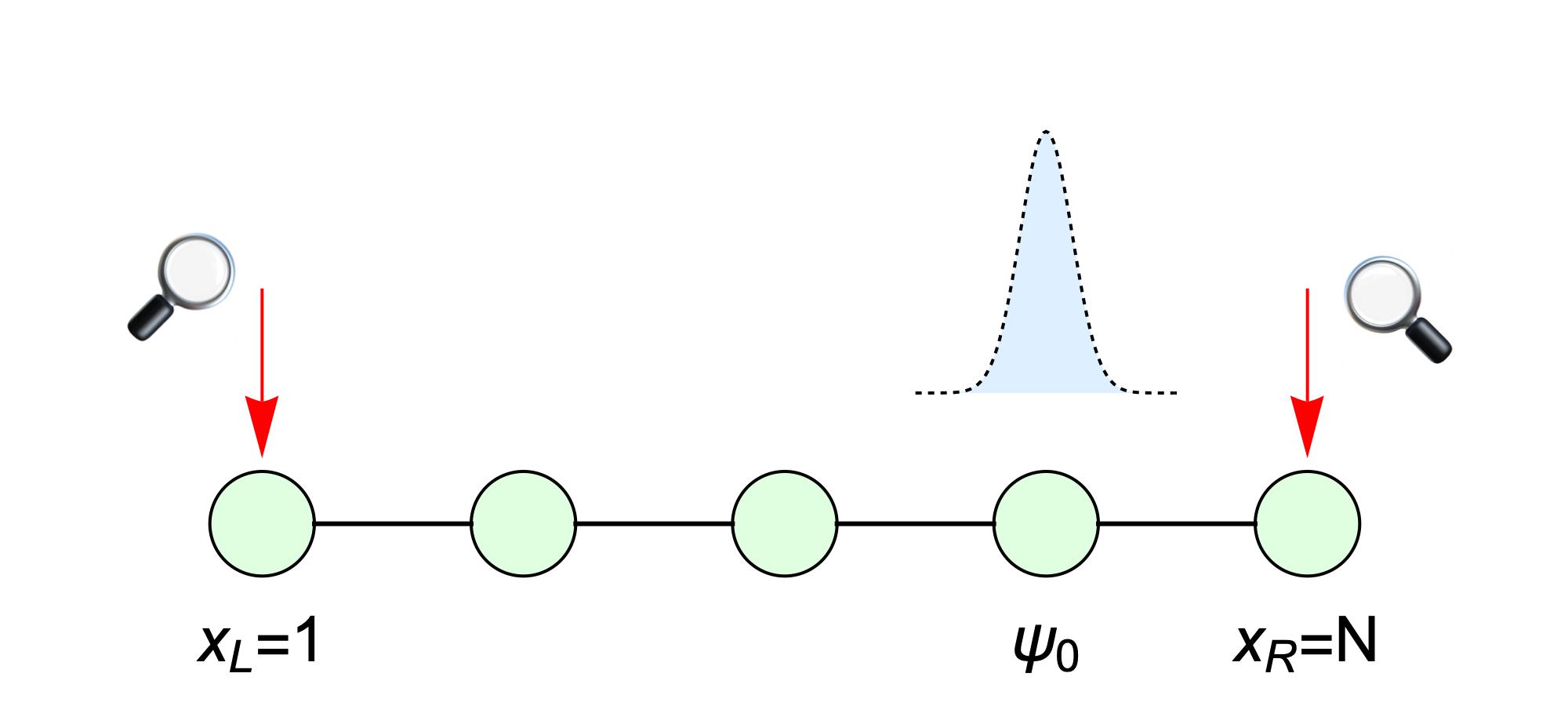} 
	\caption{A continuous-time quantum walk initially localized at $\ket{\psi_0}=\ket{4}$ moves on a tight-binding chain of size $N=5$, with Hamiltonain in Eq.~\eqref{letter-main-eqq-1}. Projective measurements are performed at the end points of the chain, $x_L=1$ and $x_R=N$, at regular time intervals $\tau$ adopting the splitting protocol explained in the Letter. The goal is to determine the probabilities of detecting at the left and the right endpoints.}
	\label{letter-fig-scheme}
\end{figure}

\textit{General formalism:} We focus on a discrete-state, continuous-time quantum system characterized by a time-independent Hermitian Hamiltonian $H$, defined on a Hilbert space $\mathcal{H}$ with a finite number of states $N$ \cite{SRPS-1, Qcom5,MULKEN2011, VenegasAndraca2012, Godec2025}. The dynamics is monitored at two general and distinct target states, $\ket{x_L}$ and $\ket{x_R}$, at regular time intervals $\tau$, which define the sampling time of the dynamics. The target states satisfy the orthogonality condition $\braket{x_L|x_R} = 0$. The Hamiltonian could, for instance, represent the nearest-neighbor tight-binding model on a finite chain
\begin{align}
H = - \gamma \sum _{x=1}^{N-1} \Big[  \ket{x} \bra{x+1}+\ket{x+1} \bra{x} \Big],  \label{letter-main-eqq-1}
\end{align}
where $\gamma$ is the hopping amplitude and $x_L$ and $x_R$ could be the end points of the chain; see Fig.~\ref{letter-fig-scheme}. We will return to this model later to illustrate the general results.

\bluew{More generally, for a quantum walk on a finite graph described by a normalized state function $\ket{\psi(t)}$ at time $t$, an arbitrary Hamiltonian $H$ and general target states $\ket{x_L}$ and $\ket{x_R}$, the time-evolution takes place as follows - (1) Initialize the system in an arbitrary state $\ket{\psi(0)}=\ket{\psi_0}$. (2) Evolve it for a time duration $\tau$ as $\ket{\psi(\tau)} = U(\tau ) \ket{\psi_0} $ with the unitary operator $U(\tau) \equiv \exp (-iH \tau)$, where $\hbar=1$. (3) At the end of this time, perform a projective measurement to check if the system is in the $\ket{x_L}$ target state. If detected, the process stops. Otherwise, the state $\ket{\psi(\tau)}$ is projected onto the subspace complementary to $\ket{x_L}$, yielding $\ket{\psi(\tau^+)}=\mathcal{N}\left(\mathbbm{1}-\ket{x_L}\bra{x_L}\right)\ket{\psi(\tau)} $ \cite{Cohenbook}. Here $\tau^+$ denotes the instant immediately after the measurement, $\mathbbm{1}$ is the identity operator, and $\mathcal{N}$ ensures normalization $\braket{\psi(\tau^+)|\psi(\tau^+)}=1$. (4) With $\ket{\psi(\tau^+)}$, a second projective measurement is immediately performed onto the second target state $\ket{x_R}$. If detected there, the process terminates. Otherwise, the state $\ket{\psi(\tau^+)}$ is again projected onto the subspace complementary to $\ket{x_R}$ similarly as before. (5) Starting from this new projected state, repeat steps~(2-4).}

We are interested in the splitting probability, \emph{i.e.}, the probability that the system is eventually detected in one target before the other one. We therefore repeat steps~(2-4), in principle, for an infinite number of times or until a successful detection occurs at a target. Notice that after every measurement cycle, a projective measurement is first performed at $\ket{x_L}$ and subsequently at $\ket{x_R}$. However, our results are insensitive to this ordering, and one may equally reverse the sequence without affecting any of the conclusions \cite{SinghSM}.




\textit{Splitting probability:} Suppose that the measurement cycle is repeated $n$ times (with $n \geq 1$). The probability that the system is first detected at time $t=n\tau$ at the $\ket{x_{\alpha}}$ target (with $\alpha\in \{L,R \}$), before any detection occurs at the other target state $\ket{x_{\overline{\alpha}}}$ (where $\overline{\alpha}=R$ if $\alpha=L$ and vice versa), is given by $F_n^{(\alpha)} = \big|\phi_n^{(\alpha)}  \big|^2$, where \cite{SMSingh2}
\begin{equation}
\scalebox{1.0}{$
\begin{split}
\phi _n^{(\alpha)} =  \bra{x_{\alpha}} U(\tau) \mathcal{S}^{n-1}\ket{ \psi _0}.
\end{split}$} \label{letter-unitary-eqn-sa11} 
\end{equation}
Here, $\mathcal{S} \equiv \Big[ \Big(\mathbbm{1}-\ket{x_L}  \bra{x_L} - \ket{x_R}  \bra{x_R} \Big) U(\tau) \Big]$ is the so-called survival operator \cite{QW6, QW7,QW8}. It demonstrates unitary evolution in the time interval $\tau$, followed by the complementary projection $\Big(\mathbbm{1}-\ket{x_L}  \bra{x_L} - \ket{x_R}  \bra{x_R} \Big)$, indicating failed detections of both target states. 
In Eq.~\eqref{letter-unitary-eqn-sa11}, the survival operator appears $(n-1)$ number of times describing the $(n-1)$ consecutive failed detections of the targets. During the final interval from $t=(n-1)\tau$ to $t=n\tau$, the system evolves unitarily, giving rise to the factor $U(\tau)$ in Eq.~\eqref{letter-unitary-eqn-sa11}. We then perform the $n$-th measurement sequentially, first probing $\ket{x_L}$ and, if no detection occurs, projecting with $(\mathbbm{1}-\ket{x_L}\bra{x_L})$ before subsequently probing $\ket{x_R}$. Since $\braket{x_R|x_L}=0$, this intermediate projection does not affect the amplitude. \bluew{Accounting for the probability of failed detections in all $(n-1)$ previous measurements, we then obtain Eq.~\eqref{letter-unitary-eqn-sa11} for both $\ket{x_L}$ and $\ket{x_R}$ \cite{SMSingh2}.}

The quantity $\phi _n^{(\alpha)}$ thus can be interpreted as the first-detection amplitude of the $\ket{x_{\alpha}}$ state at the $n$-th measurement attempt before the other $\ket{x_{\overline{\alpha}}}$ state is detected. The corresponding splitting probability of detection is $P_{\alpha}(\psi_0) = \sum _{n=1}^{\infty} \big| \phi_{n}^{(\alpha)}\big|^2 $. The sum accounts for the events of first detection occurring at the first, second, third measurements, and so on. For notational simplicity, we suppress the $N$ and $\tau$ dependence in $P_{\alpha}(\psi_0)$. \bluew{Note that $P_L(\psi_0)+P_R(\psi_0)$ need not equal unity, since dark subspaces can prevent detection at either target, as we show later. Thus, $\phi_n^{(\alpha)}$ should be understood as a first-detection amplitude, rather than as the amplitude of a normalized Hilbert-space state.}

\textit{Mapping to single-target problems:} To enable a systematic comparison with the classical splitting probabilities in Eq.~\eqref{lett-unitary-eqn-classical}, we focus on the one-dimensional chain with $N$ sites. We label the sites as $x \in \{1,2, \ldots N \}$. Arrival is monitored at the endpoints, $x_L=1$ and $x_R=N$, of the chain. We now show that $\phi_n^{(\alpha)}$ admits an exact mapping to a pair of single-target problems for a broad class of Hamiltonians. 

To this end, we introduce two auxiliary orthogonal quantum states $\ket{d_{\pm}} \equiv \left( \ket{x_L} \pm \ket{x_R} \right)/\sqrt{2}$. For $n=1$, one can express Eq.~\eqref{letter-unitary-eqn-sa11} in terms of these states as $\phi_1^{(L)} =\left[ \braket{d_+|U(\tau)|\psi_0} + \braket{d_-|U(\tau)|\psi_0} \right] /\sqrt{2}$, and similarly for the right boundary. Notice that $\phi_1^{(L)}$ decomposes into two separate terms $\braket{d_{\pm}|U(\tau)|\psi_0}$. Clearly, these terms are single-target amplitudes of detecting the system in states $\ket{d_{\pm}}$ in the first measurement \cite{QW6, QW7,QW8}.

For $n=2$, the decomposition becomes more involved, \small{$\phi_2^{(L)} = [ \braket{d_+|U(\tau) \left(\mathbbm{1}-\ket{d_+}\bra{d_+} \right)U(\tau)|\psi_0}+\braket{d_-|U(\tau) \left(\mathbbm{1}-\ket{d_-}\bra{d_-} \right)U(\tau)|\psi_0} - \mathcal{M}] / \sqrt{2}$, where $\mathcal{M} = \braket{d_+|U(\tau)|d_-} \braket{d_-|U(\tau)|\psi_0} +\braket{d_-|U(\tau)|d_+} \braket{d_+|U(\tau)|\psi_0}$}. \normalsize{Notice} that $\mathcal{M}$ involves matrix elements such as $\braket{d_{\pm}|U(\tau)|d_{\mp}}$. For a generic Hamiltonian, these elements are generally nonzero, and consequently $\mathcal{M}$ does not vanish. However, for parity-symmetric Hamiltonians, we show in the Supplemental Material (SM) that $\braket{d_{\pm}|U(\tau)|d_{\mp}}=0$ \cite{SMSingh2}. The parity operator is defined as $\mathcal{P}\ket{x}=\ket{N+1-x}$, corresponding to reflection about the center of the chain. For Hamiltonians satisfying this symmetry (\emph{i.e.,} the commutation $[H, \mathcal{P}] = 0$), one gets $\mathcal{M}=0$, and the above expression of $\phi_2^{(L)}$ reduces to a linear combination of two single-target detection amplitudes at $\ket{d_{\pm}}$ for $n=2$.

In the SM \cite{SMSingh2}, we use induction to show that such a decomposition is valid for generic $n$
\begin{equation}
\scalebox{0.9}{$
\begin{split}
&\phi _{n}^{\left(  L \right)}  = \left( \chi _{n}^{\left(+\right)}  + \chi _{n}^{(-)} \right) \big/ \sqrt{2},~~\phi _{n}^{\left(  R \right)}  = \left( \chi _{n}^{\left(+\right)}  - \chi _{n}^{(-)} \right) \big/ \sqrt{2}, \\
&~~\text{where } \chi _{n}^{\left( \pm \right)}=\braket{d_{\pm} |U(\tau) \left[ \left(\mathbbm{1}-\ket{d_{\pm}} \bra{d_{\pm}} \right) U(\tau)\right]^{n-1}|\psi_0}.
\end{split}$} \label{letter-neq-1} 
\end{equation}
The terms $\chi _{n}^{\left( \pm \right)}$ represent the first-detection amplitudes for two auxiliary
quantum systems, each stroboscopically monitored at a distinct single target state \cite{QW6, QW7,QW8}. For $\chi _{n}^{\left( + \right)}$, the auxiliary target state is $\ket{d_+}$, whereas for $\chi _{n}^{\left( - \right)}$, the dynamics is monitored at a different auxiliary target state $\ket{d_-}$. Eq.~\eqref{letter-neq-1} is the mapping to a pair of single-target detection problems, as mentioned before. 


The mapping is a direct consequence of the superposition principle, which permits the introduction of the quantum states $\ket{d_{\pm}}$. Such a construction has no classical analogue, since $\ket{d_{\pm}}$ are not physical states in the classical probabilistic framework. Importantly, Eq.~\eqref{letter-neq-1} provides a practical advantage: it enables us to use earlier methods from the single-target first detection setups \cite{QW6, QW12,QW16, QW21, QW7, QW8,QW13} to obtain explicit formulas for the splitting probabilities, as we demonstrate later. From a physical perspective, the mapping implies an interference effect between the two single-target detection processes, which can be either constructive or destructive. As we soon show, this interference leads to a phase-transition-like behavior in the splitting probabilities.\\


\begin{figure}[t]
	\centering
	\includegraphics[scale=0.25]{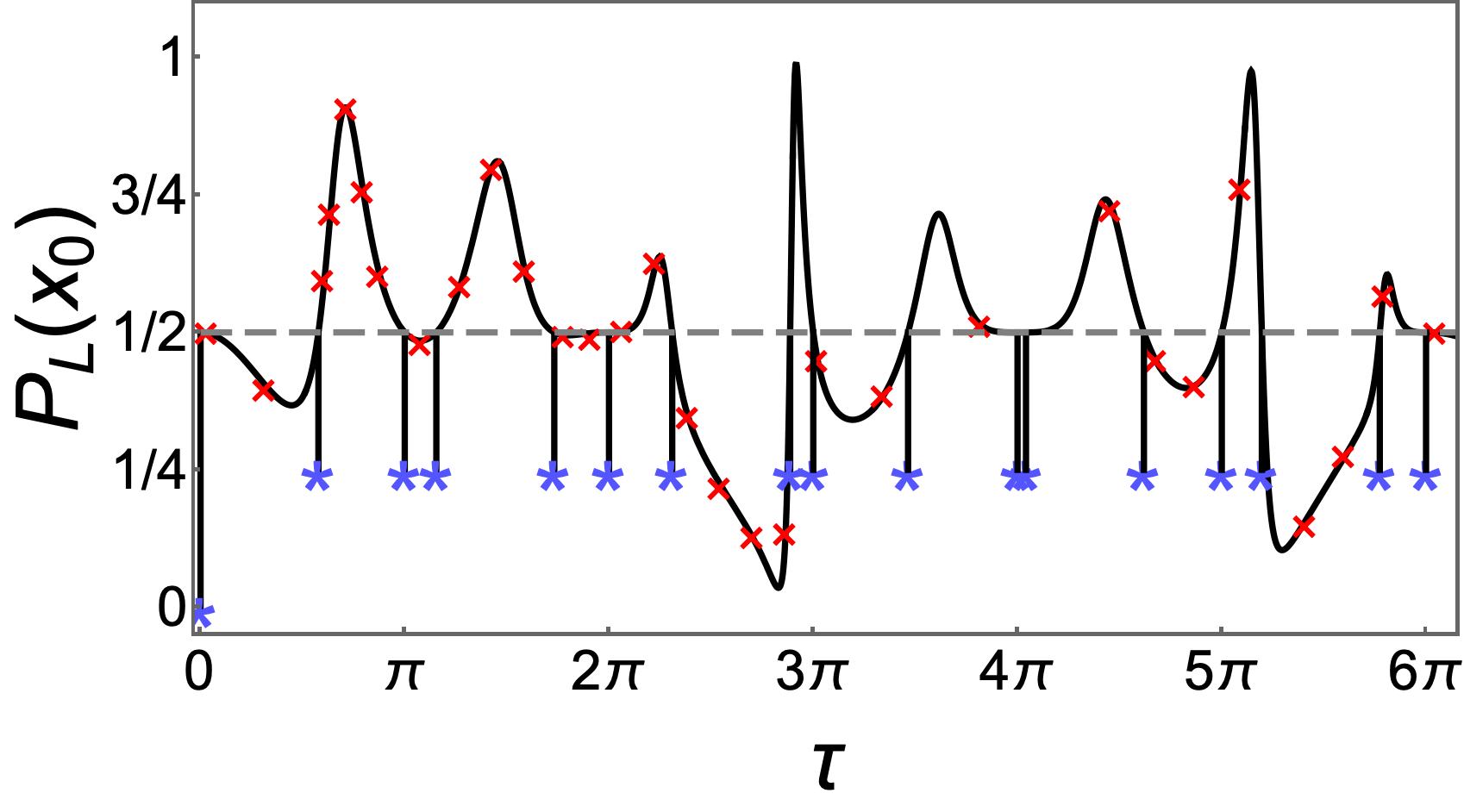} 
	\caption{Splitting probability at the left boundary as a function of $\tau$ for $N=5$ and $x_0=4$ (\emph{i.e.}, $\ket{\psi_0}=\ket{4}$). Red and blue symbols represent the numerical simulations with blue highlighting the resonant points. Black solid line corresponds to our analytical results in Eqs.~\eqref{letter-res-cond} and \eqref{letter-neq-2}. When $P_L(4)>1/2$ (shown by the dashed line), the classical proximity effect is broken, as explained in the Letter.}
	\label{letter-fig-resonance}
\end{figure}
\textit{Tight-binding model with nearest-neighbor hops:} For this case, the Hamiltonian takes the form in Eq.~\eqref{letter-main-eqq-1}. Observe that $H$ is parity symmetric, \emph{i.e.}, $[H,\mathcal{P}]=0$. Moreover, this Hamiltonian has been also implemented experimentally \cite{ExptQC-1, ExptQC-2, ExptQC-3, ExptWG-1, ExptWG-3, Tang2018, ExptWG-2, Liuexpt1}.
Throughout, we rescale time by $\gamma t/\hbar \to t$ or equivalently, we set $\gamma=\hbar=1$. Then, the dispersion relation becomes $E_k = -2 \cos \left( \omega _k \right)$ and the normalized energy eigenfunctions $\psi_k(x)=\braket{x|E_k} = \sqrt{2/(N+1)} \sin \left( \omega_k x \right) $, where $\omega _k = \pi k /(N+1)$, $k \in \{1,2, \ldots, N \}$ and $x \in \{1,2, \ldots, N \}$. For simplicity, we take the initial state $\ket{\psi_0} = \ket{x_0}$, where recall $x_0 \in \{1,2, \ldots N \}$, $x_L=1$ and $x_R=N$.

\textit{Dark States and the Splitting problem:} Employing our mapping~\eqref{letter-neq-1}, we find that at certain values of $\tau$, both $P_L(x_0)$ and $P_R(x_0)$ develop pointwise discontinuities (see the blue symbols in Fig.~\ref{letter-fig-resonance}). At these points, the total detection probability drops below unity, $P_L(x_0)+P_R(x_0)<1$. A finite fraction of realizations of the walk never triggers a detection at either boundary. Contrast this behavior with Eq.~\eqref{lett-unitary-eqn-classical} where the sum is always equal to unity, since the underlying classical process is ergodic. 
\bluew{In the SM \cite{SMSingh2}, we prove that the discontinuities arise whenever the sampling time satisfies}
\begin{equation}
\left(E_k - E_\ell\right)\tau = 0  \text{ mod }  2 \pi,
\label{letter-res-cond}
\end{equation}
for any two distinct energies, $E_{k}$ and $E_{\ell}$,
such that the corresponding energy eigenstates, $\ket{E_k}$ and $\ket{E_{\ell}}$, have the same parity under the reflection transformation introduced before. Resonances do not occur if their parities are different. \bluew{The parity-specific condition follows from Eq.~\eqref{letter-neq-1}, which maps the overall splitting problem into two parity-decomposed auxiliary single-target detection problems \cite{SMSingh2}.}

Physically, at special $\tau$ in Eq.~\eqref{letter-res-cond}, a dark state $\ket{\mathcal{D}} \sim \left[ \sin(\omega_k) \ket{E_{\ell}}-\sin(\omega_{\ell}) \ket{E_{k}} \right]$ is created in the system, that is orthogonal to both boundaries, $\braket{x_L|\mathcal{D}} = \braket{x_R|\mathcal{D}} = 0$, at the measurement instances. The process of repeated measurements can drive the system to this state, thus creating a non-detectable state and hence the splitting probabilities sum is less than unity. More specifically, if the initial state has a finite overlap with this state, $\braket{\mathcal{D}|x_0} \neq 0$, then Eq.~\eqref{letter-res-cond} implies that the system evolves in such a way that a component of its wave function becomes orthogonal to both boundaries at the measurement times $t=\tau,~2 \tau,~3 \tau, \ldots$ \cite{SinghSM}. This component of the wave function is never detected at the boundaries and thus the total detection probability is less than one, as mentioned above. In Fig.~\ref{letter-fig-resonance}, we verify the resonances using numerical simulations for $x_0 = 4$ and $N = 5$ (see the End Matter for details).

\textit{Loss of proximity effect:} For generic, non-resonant values of $\tau$, the splitting probabilities depend smoothly on $\tau$, as shown by the red symbols in Fig.~\ref{letter-fig-resonance}, and satisfy the normalization condition $P_L(x_0)+P_R(x_0)=1$. Their expressions are
\begin{equation}
\scalebox{0.95}{$
\begin{split}
P_L(x _0) = \frac{1}{2}+ \xi (x _0,N, \tau),~ P_R(x _0) = \frac{1}{2}-\xi (x _0,N, \tau),
\end{split}$} \label{letter-neq-2}
\end{equation}
where the function $ \xi (x _0,N, \tau)$ represents the interference effect between  the single-target detection amplitudes $\chi_{n}^{(\pm)}$ in Eq.~\eqref{letter-neq-1}. It can be computed exactly and is provided in Eq.~\eqref{SI-4} at the End Matter. For example, when $x_0=4$ and $N=5$, we get
\begin{equation}
\scalebox{0.85}{$
\begin{split}
\xi (4,5, \tau) = -\frac{\sqrt{3} \sin \left(\frac{\tau}{2} \right)^2 \sin \left( \tau \right) \sin \left( \sqrt{3}  \tau \right) }{4+ \cos \left( 2  \tau \right)  + \cos \left(  \tau \right) \left[  -1+ \cos \left( \sqrt{3 } \tau \right) \left\{ -5+  \cos \left(  \tau \right) \right\} \right]}.
\end{split}$} \label{ajniuyqb}
\end{equation}
As shown in Fig.~\ref{letter-fig-resonance}, over a wide range of $\tau$ values, we find $P_L(4) > 1/2$, and hence $P_R(4)<1/2$ according to Eq.~\eqref{letter-neq-2}. Thus, the proximity effect is lost, $P_R(4) < P_L(4)$, even though the walker $(x_0=4)$ is initially closer to the right boundary $(x_R=5)$. This behavior contrasts with the classical case in Eq.~\eqref{lett-unitary-eqn-classical}.

\begin{figure}[t]
	\centering
	\includegraphics[scale=0.25]{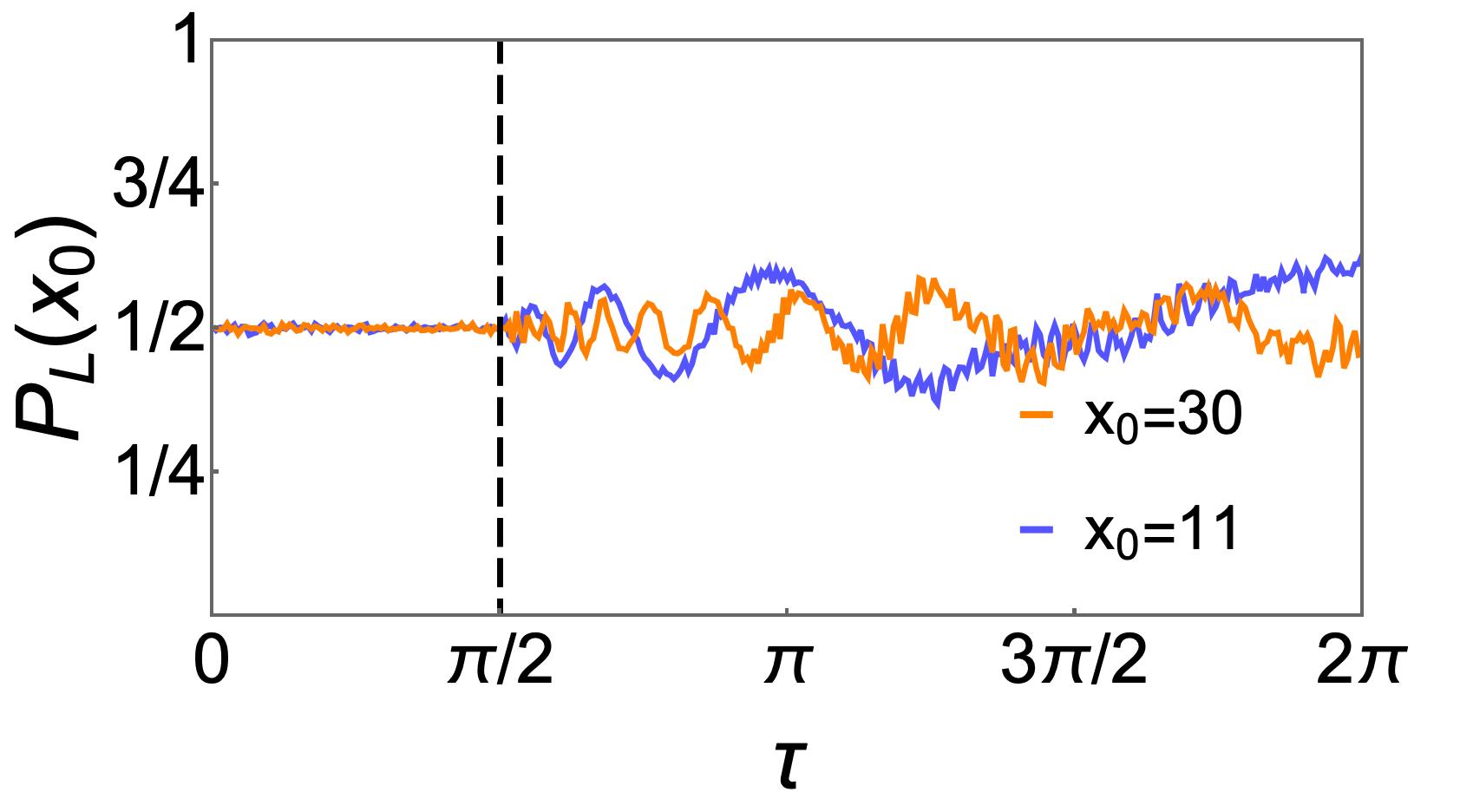} 
\caption{Flat-to-fluctuating transition of the splitting probability at $\tau _c = \pi/2$ for two initial conditions, $x_0=11$ and $x_0=30$ with $N=400$ using Eq.~\eqref{letter-neq-2}. Resonances at special values of $\tau$, given in Eq.~\eqref{letter-res-cond}, are omitted for clarity.}
\label{letter-fig-transition}
\end{figure}

\textit{Flat-to-fluctuating transition for large system size:} We now turn to the sampling transition that was mentioned before. Fig.~\ref{letter-fig-transition} shows the plot of $P_L(x _0)$ as a function of $\tau$ using Eq.~\eqref{letter-neq-2} for two initial conditions, $x_0=11$ and $x_0=30$, keeping the system size fixed at $N=400$ for both. For $N \gg 1$, the splitting probability undergoes a surprising change in behavior at a critical sampling time $\tau_c = \pi/2$, shown by the dashed vertical line in Fig.~\ref{letter-fig-transition}. This critical value, as shown later, is related to the width of the energy band. In the regime $0< \tau \leq \tau _c$, the splitting probability is universal, essentially equal to $1/2$, independent of both $\tau$ and $x_0$. 
For $\tau > \tau_c$, a qualitatively different, non-universal behavior emerges. Here, $P_L(x _0)$ deviates from $1/2$, and develops a fluctuating pattern with pronounced peaks and dips as the sampling time is varied. These features are not universal and depend on both $x_0$ and $\tau$. Thus, as $\tau$ crosses $\tau_c$, $P_L(x_0)$ changes from a universal flat behavior to a fluctuating, non-universal behavior. \bluew{The same behavior is observed for other values of $N$ \cite{SMSingh2} as well as for $P_R(x_0)$ \cite{SinghSM}.} 
To gain some insight into this behavior, we start with an analysis of the Zeno limit.


\textit{Small-$\tau$ analysis:} When $\tau$ is small but finite $(0 < \tau \ll 1)$, we can perform a perturbative analysis in $\tau$ to obtain $\xi(x_0,N, \tau) \simeq  \left( \delta _{x_0,1}-\delta_{x_0,N}  \right)/2$ for general $N$ \cite{SinghSM}. Plugging this result in Eq.~\eqref{letter-neq-2} yields $P_L(x_0) \simeq P_R(x_0) \simeq 1/2$ for $1 < x_0 <N$. On the boundaries, we have $P_L(x_0=1) \simeq 1$ and $P_L(x_0=N) \simeq 0$ (and similarly for $P_R (x_0)$) as $\tau \to 0$, since for $x_0=1$, the particle is always detected at the left boundary, whereas for $x_0=N$, it is always detected at the right boundary and never at the left boundary. For all other $x_0$, the splitting probabilities take on a flat value of $1/2$. This perturbative result is consistent with the flat behavior observed in the $0<\tau\le\tau_c$ regime in Fig.~\ref{letter-fig-transition}. 

Our Zeno result is in stark contrast with the classical case in Eq.~\eqref{lett-unitary-eqn-classical}, where splitting probabilities depend linearly on $x_0$. Physically, the flatness stems because, for small $\tau$, the quantum walker is mostly confined in the bulk, with only a small probability $(\sim \tau^2)$ of being detected at the boundaries during each measurement attempt. As a result, it typically requires many attempts before being detected, and over this long sequence of measurements, any memory of the initial positional asymmetry is effectively erased. This explains the flatness of the splitting probability at small $\tau$, yet does not explain the transition at $\tau_c$.


\begin{figure}[t]
	\centering
	\includegraphics[scale=0.11]{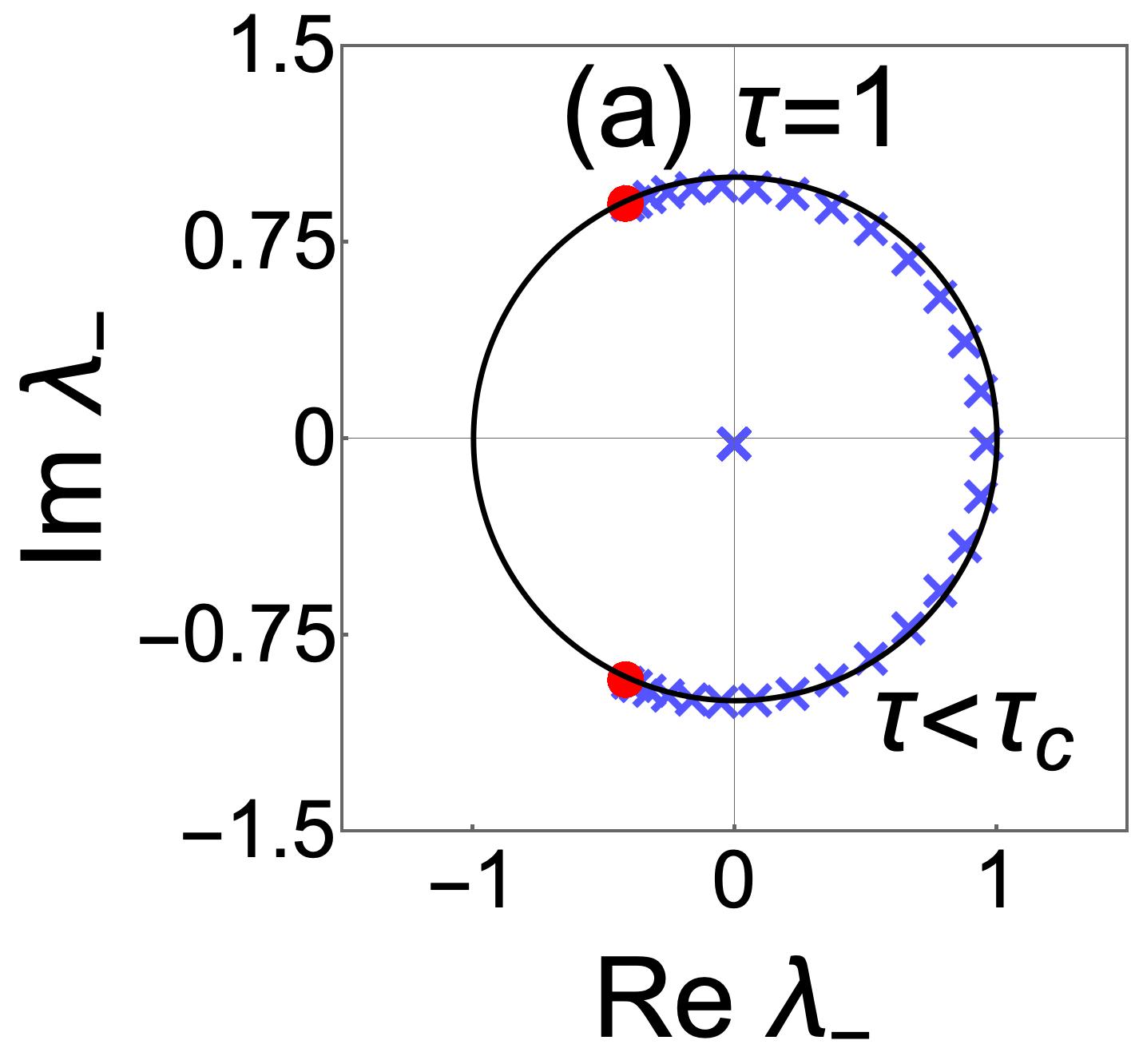} 
	\includegraphics[scale=0.11]{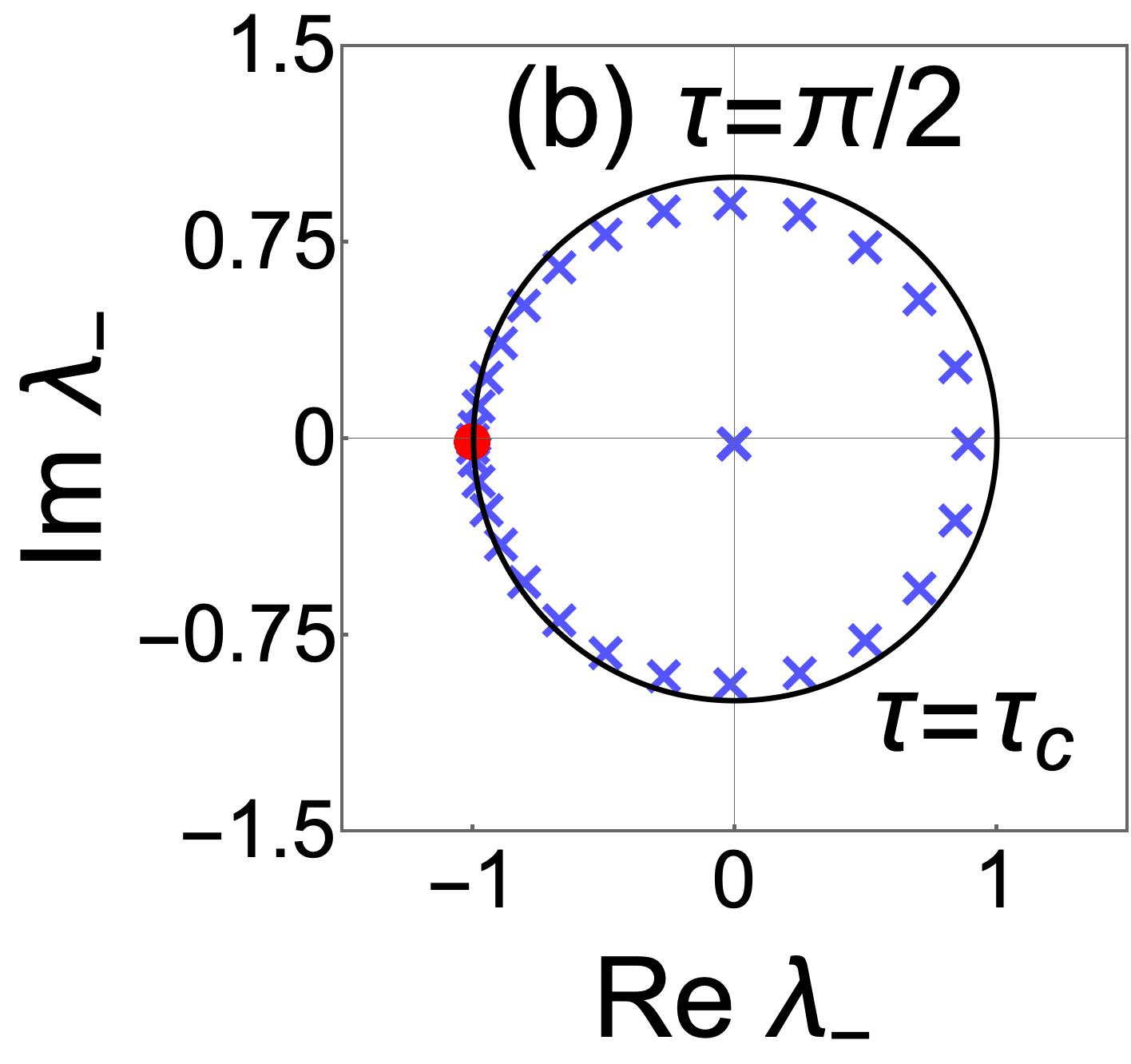} 
	\includegraphics[scale=0.11]{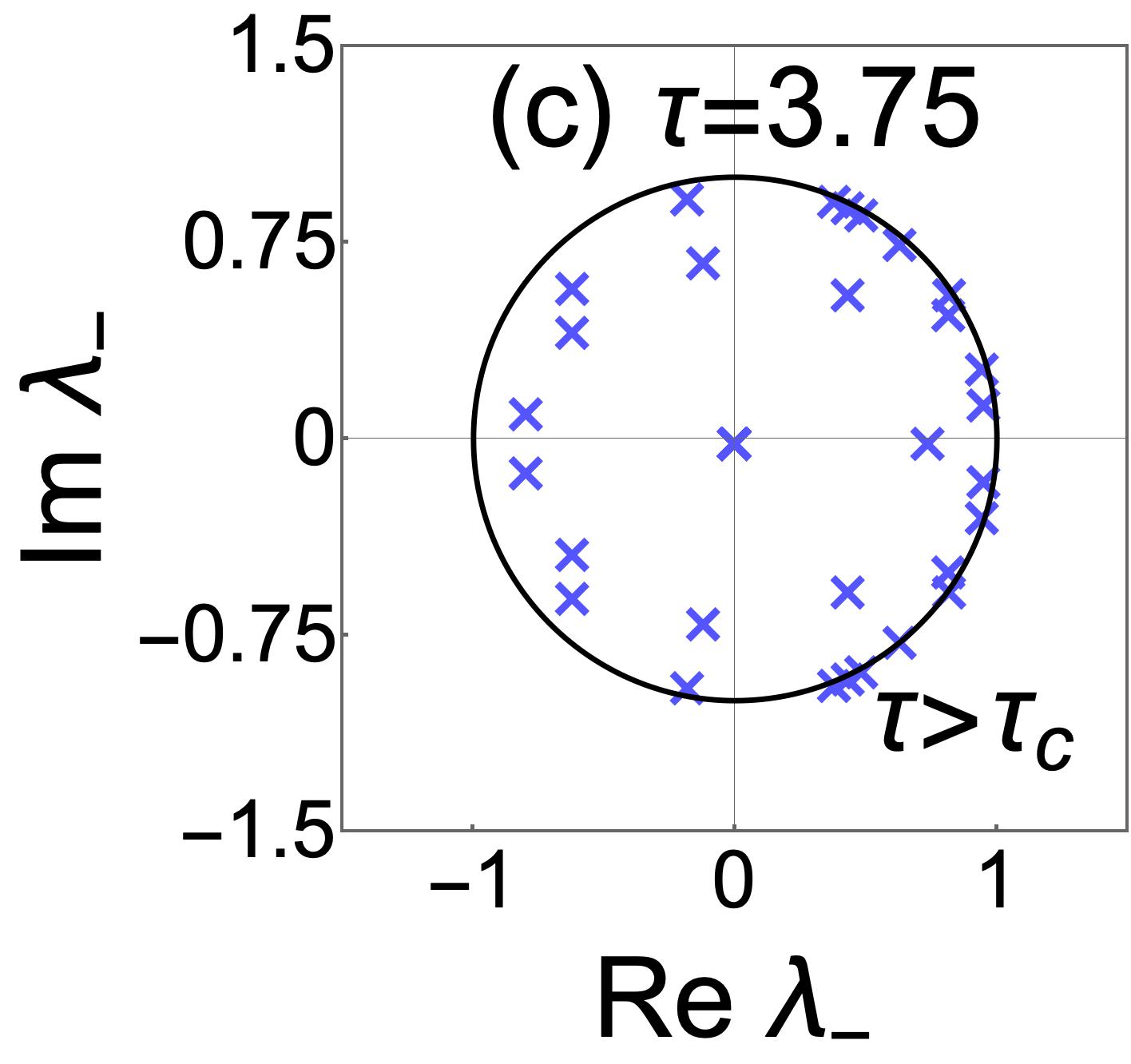} 
\caption{Transition in the arrangement of the eigenvalues $\{ \lambda_{-} \}$ (in blue) of the survival operator $\mathcal{S}_- =  \left( \mathbbm{1}-\ket{d_{-}}\bra{d_{-}} \right)U(\tau)$ in Eq.~\eqref{letter-neq-1} as we go from (a) $\tau =1~(\tau < \tau_c)$, (b) $\tau = \pi/2~(\tau = \tau_c)$ and (c) $\tau = 3.75~(\tau > \tau_c)$ with $N=61$. In (a-b), the two red points represent $a_{1}(\tau) =\exp\!\left(-iE_{\min}\tau\right)$ and $a_2(\tau)=\exp\!\left(-iE_{\max}\tau\right)$, as explained in the Letter. The transition in the arrangement of $\{ \lambda_-\}$ manifests as the transition in the splitting probability. For $x_0=11$, we find $\xi(x_0, N, \tau=1) \approx 0.002$, $\xi(x_0, N, \tau=\tau_c) \approx -0.002$ and $\xi(x_0, N, \tau=3.75) \approx -0.135$.}
\label{letter-fig-pole}
\end{figure}
\textit{Spectral origin of the transition:} \bluew{Eq.~\eqref{letter-neq-2} shows that deviations of the splitting probabilities from $1/2$ originate from the term $\xi(x_0,N, \tau)$. As mentioned before, this term represents an interference effect between $\chi_n^{(\pm)}$. However, it can nevertheless be expressed in terms of the eigenvalues of either $\mathcal{S}_{-} \equiv (\mathbbm{1}-\ket{d_{-}}\bra{d_{-}})U(\tau)$ or $\mathcal{S}_{+}\equiv (\mathbbm{1}-\ket{d_{+}}\bra{d_{+}})U(\tau)$. Here $\mathcal{S}_{\pm}$ are the survival operators in the auxiliary single-target detection problems in Eq.~\eqref{letter-neq-1}. Below we explain the flat-to-fluctuating transition using $\mathcal{S}_{-}$ and relegate the corresponding discussion using $\mathcal{S}_{+}$ to the SM~\cite{SMSingh2}.}


Let $\{\lambda_-\}$ denote the eigenvalues of $\mathcal{S}_-$.  Each $\lambda_-$ yields an individual contribution, and $\xi(x_0,N,\tau)$ is determined by their combined sum.
For $\tau \leq \tau _c$ and large system sizes, we find in Fig.~\ref{letter-fig-pole}(a) (in blue) that all $\{ \lambda_{-} \}$, except the one at the origin, are arranged regularly along a single arc close to $|z|=1$, between the points $a_{1}(\tau)$ and $a_{2}(\tau)$ (in red). By inspection, we find $a_{1}(\tau) =\exp\!\left(-iE_{\min}\tau\right)$ and $a_2(\tau)=\exp\!\left(-iE_{\max}\tau\right)$, where $E_{\min} \approx -2$ and $E_{\max} \approx 2$ are the minimum and maximum eigenvalues of $H$ in Eq.~\eqref{letter-main-eqq-1}. Because of the regular arrangement of $\{ \lambda_{-} \}$ in the complex plane, their individual contributions to $\xi(x_0,N,\tau)$ largely cancel with each other, yielding a negligible sum, \emph{i.e.} $\xi(x_0,N \gg 1 , \tau \leq \tau_c) \approx 0$. Eq.~\eqref{letter-neq-2} then implies $P_L(x_0) \approx P_R(x_0) \approx 1/2$.

This behavior persists till the critical sampling time
\begin{equation}
    \Delta E~ \tau_c = 2 \pi,
\end{equation}
with $\Delta E = \left( E_{\max} -E_{\min} \right)= 4$ (\emph{i.e.}, $\tau_c=\pi/2$ as pointed out before), the two endpoints meet, $a_1(\tau)=a_2(\tau)=-1$, and the arc closes (Fig.~\ref{letter-fig-pole}(b)). 

Increasing $\tau$ further, the points $a_1(\tau)$ and $a_2(\tau)$ move past each other, and $\{ \lambda_{-} \}$ are no longer confined to a single arc; see Fig.~\ref{letter-fig-pole}(c). Instead, their arrangement becomes irregular in the complex plane, with eigenvalues spreading nonuniformly inside the unit disk. As a result, the contributions associated with different $\lambda_-$ no longer cancel, and $\xi(x_0,N \gg 1,\tau > \tau_c) \neq 0$. This leads to the departure of the splitting probabilities from their flat value of $1/2$ via Eq.~\eqref{letter-neq-2} for $\tau > \tau_c$. In this way, the transition in the splitting probabilities is directly linked to the qualitative reorganization of the eigenvalues $\{ \lambda_- \}$ as $\tau$ crosses $\tau_c$; see \cite{SMSingh2} for further details.

\bluew{\textit{Hadamard walks:} Tuning the sampling time in monitored continuous-time quantum walk with two targets therefore leads to a variety of interesting, nonclassical effects. As mentioned in the Introduction, previous works considered the discrete-time Hadamard walk \cite{Vishwanath2001, SP1, Bach2009, SP5, Ammara2025AbsorbingBoundaries}. This model features a quantum particle moving on a one-dimensional chain with two absorbing boundaries with the dynamics being governed by an internal two-state coin degree of freedom. At every time step the coin undergoes the Hadamard transformation, the particle moves conditioned on the coin state, and the measurement is performed at the same time step. Our continuous-time quantum walk differs from this setting in two respects. Firstly, there is no internal coin degree of freedom in our setup, and a successful measurement specifies the full state of the walker. By contrast, the Hadamard-walk approach employs subspace measurements \cite{QW22, Liu2024}, where only the position is monitored while the coin state remains unresolved. Secondly, in the Hadamard-walk studies \cite{Vishwanath2001, SP1, Bach2009, SP5, Ammara2025AbsorbingBoundaries}, the system evolution timescale and the sampling timescale are the same. In our setting, by contrast, tuning the sampling time relative to the intrinsic system timescale is essential for observing the flat-to-fluctuating transition and the pointwise discontinuities  shown in blue in Fig.~\ref{letter-fig-resonance}. To the best of our knowledge, these phenomena have not been observed so far in Hadamard walks.}

\textit{Experimental outlook:} Our results can be tested experimentally. In quantum computers \cite{ExptQC-1, ExptQC-2, ExptQC-3}, for instance, monitored quantum walks can be implemented using qubits, and mid-circuit measurements enable periodic measurement on the target states. Experimental implementations are also possible in photonic waveguides  \cite{ExptWG-1, ExptWG-3, Tang2018, ExptWG-2, Liuexpt1} and trapped-ion platforms with periodic laser pulses \cite{Ryan2025}. These setups provide direct experimental platforms to probe the splitting dynamics and the flat-to-fluctuating transition predicted here. Notably, observing this transition does not require very large system sizes, and the effect already appears for moderate values of $N$ \cite{SinghSM}, making it amenable for experiments.

\textit{Conclusion and outlook:}  We have shown that the splitting probability of a continuous-time quantum walk exhibits a sharp transition from a universal regime where it is equal to $1/2$, independent of the initial position and the sampling time to a non-universal regime, where it has fluctuating structures. 
Notably, the critical point $\tau _c$ is dependent only on the width of the energy band. Therefore, the transition is likely to go beyond the studied model and to hold generically for a wide class of Hamiltonians. Other features include the breakdown of the proximity effect and the emergence of discontinuous jumps at special sampling times. \bluew{All these features are in stark contrast to the classical case in Eq.~\eqref{lett-unitary-eqn-classical}. Even when the classical random walk is monitored stroboscopically, these features do not arise, as discussed in the End Matter as well as in the SM \cite{SMSingh2}.}

\bluew{Our results are derived for a finite chain. However, it is possible to extend our splitting framework to an infinite lattice with two monitored target sites. Then the walker can leave the interval between the two targets and never return \cite{QW21, QW6, QW12}. It would be interesting to explore how the absence of recurrence modifies the splitting probability. Extending the splitting problem to many-body systems is another natural future direction \cite{Gambassi2025,Liu2024}. Since recent work on recurrence to a single target state has already revealed two critical sampling timescales and dynamical transitions in the return probability \cite{Gambassi2025}, the dual-target case is expected to display a rich structure.} 

\textit{Acknowledgment:} This research was supported by the Israel Science Foundation (grant No. 2311/25).

\begin{widetext}

\begin{center}
\textbf{End Matter}
\end{center}
\setcounter{equation}{0}
\renewcommand{\theequation}{EM\arabic{equation}}
\renewcommand{\theHequation}{EM.\arabic{equation}}
\begin{center}
\textit{General Expressions of $P_L(x_0)$ and $P_R(x_0)$}
\end{center}
In Eq.~\eqref{letter-neq-1}, we explained that the detection amplitudes, $\phi_n^{(L)}$ and $\phi_n^{(R)}$, are mapped to two auxiliary single-target detection amplitudes $\chi _{n}^{\left( \pm \right)}=\braket{d_{\pm} |U(\tau) \left[ \left(1-\ket{d_{\pm}} \bra{d_{\pm}} \right) U(\tau)\right]^{n-1}|x_0}$. Summing the squared moduli of the first-detection amplitudes over $n$ yields the splitting probabilities
\begin{align}
& P_L(x_0) = \sum _{n=1}^{\infty} \big| \phi_n^{(L)} \big|^2 = \frac{1}{2} \left[ \rho _+(x_0) + \rho _-(x_0) \right]+  ~\text{Re} \left[ \sum_{n=1}^{\infty} \chi _{n}^{\left( + \right)} ~\left( \chi _{n}^{\left( - \right)} \right)^* \right], \label{SI-1} \\
& P_R(x_0) = \sum _{n=1}^{\infty} \big| \phi_n^{(R)} \big|^2 = \frac{1}{2} \left[ \rho _+(x_0) + \rho _-(x_0) \right]- ~\text{Re} \left[ \sum_{n=1}^{\infty} \chi _{n}^{\left( + \right)} ~\left( \chi _{n}^{\left( - \right)} \right)^* \right]. \label{SI-2}
\end{align}
Here, $\rho _{\pm}(x_0) = \sum _{n=1}^{\infty} \big| \chi _{n}^{\left( \pm \right)}  \big|^2  $ are the total detection probabilities in the auxiliary problems. Single-target detection problems of this type have been extensively studied before \cite{QW6, QW12,QW16, QW21, QW7, QW8,QW13}. Leveraging these methods within our mapping, we 
get $\rho _{\pm}(x_0) $ that appear in Eqs.~\eqref{SI-1} and \eqref{SI-2}. For the Hamiltonian in Eq.~\eqref{letter-main-eqq-1} and non-special $\tau$, i.e., $\tau$ not satisfying Eq.~\eqref{letter-res-cond}, we find \cite{SMSingh2}
\begin{align}
\rho _+(x _0) = \sum_{\substack{k=1 \\ k \in  \,\text{odd}}}^{N} \big| \braket{E_k | x _0} \big|^2,~~\rho _-(x _0) = \sum_{\substack{k=1 \\ k \in  \,\text{even}}}^{N} \big| \braket{E_k | x _0} \big|^2. \label{aux-eq-4}
\end{align}
where $\ket{E_k}$ is the energy eigenstate. 
To obtain the interference part in Eqs.~\eqref{SI-1} and \eqref{SI-2}, it is instructive to introduce the auxiliary generating functions $ \widetilde{\chi}^{(\pm)}(z) =  \sum _{n=1}^{\infty} z^n  \chi _{n}^{\left( \pm \right)}  $. Using earlier methods \cite{QW6,QW16}, we can write them as \cite{SMSingh2}
\begin{align}
\widetilde{\chi}^{(+)}(z) =\frac{\sum _{k \in ~\rm{odd}} \sin (\omega _k) \sin (\omega_k x_0 ) \left(  \frac{z e^{2 i  \tau \cos (\omega _k)}}{1-z ~e^{2 i \tau \cos (\omega _k)}}  \right)  }{ \sqrt{2}  \sum _{k \in ~\rm{odd}} \sin ^2 (\omega _k)  \left( \frac{1}{1-z ~e^{2 i \tau \cos (\omega _k)}}  \right)      },~~\widetilde{\chi}^{(-)}(z) =\frac{\sum _{k \in ~\rm{even}} \sin (\omega _k) \sin (\omega_k x_0 ) \left( \frac{ze^{2 i  \tau \cos (\omega _k)}}{1-z ~e^{2 i \tau \cos (\omega _k)}}  \right)  }{  \sqrt{2} \sum _{k \in ~\rm{even}} \sin ^2 (\omega _k)  \left( \frac{1}{1-z ~e^{2 i  \tau \cos (\omega _k)}}  \right) }, \label{SI-3}
\end{align}
where $\omega _k = \pi k/ (N+1)$. In terms of these generating functions, the interference term in Eqs.~\eqref{SI-1} and \eqref{SI-2} can be expressed as
\begin{align}
\xi(x_0,N, \tau) = \text{Re} \left[ \sum_{n=1}^{\infty} \chi _{n}^{\left( + \right)} ~\left( \chi _{n}^{\left( - \right)} \right)^* \right] = \text{Re} \left[ \oint_{|z|=1} \frac{dz}{2 \pi i z}~ \widetilde{\chi}^{(+)}(z)~\left( \widetilde{\chi}^{(-)}(1/z)\right) ^* \right]. \label{ani17}
\end{align}
This relation makes it explicit that $\xi(x_0,N, \tau)$ represents the interference effect between the two single-target detection processes. Plugging Eq.~\eqref{SI-3}
\begin{align}
& \xi(x_0,N, \tau) = \text{Re} \left[ \oint_{|z|=1} dz~\mathcal{I}(z, x_0,N,\tau) \right], ~~\text{with} \label{SI-4}\\
& \mathcal{I}(z,x_0,N, \tau) = \frac{1}{4 \pi  i z}  \left[ \frac{\sum _{k \in ~\rm{odd}} \sin (\omega _k) \sin (\omega_k x_0 ) \left( \frac{e^{2 i  \tau \cos (\omega _k)}}{1-z ~e^{2 i \tau \cos (\omega _k)}}  \right)  }{   \sum _{k \in ~\rm{odd}} \sin ^2 (\omega _k)  \left( \frac{1}{1-z ~e^{2 i  \tau \cos (\omega _k)}}  \right)      } \right]  \left[ \frac{\sum _{k \in ~\rm{even}} \sin (\omega _k) \sin (\omega_k x_0 ) \left( \frac{e^{-2 i  \tau \cos (\omega _k)}}{z- ~e^{-2 i  \tau \cos (\omega _k)}}  \right)  }{   \sum _{k \in ~\rm{even}} \sin^2 (\omega _k)  \left( \frac{1}{z- ~e^{-2 i  \tau \cos (\omega _k)}}  \right)      } \right]. \label{pole-eq-2}
\end{align}
Combining Eqs.~\eqref{aux-eq-4} and \eqref{SI-4}, we obtain
\begin{align}
P_L(x_0) = \frac{1}{2} + \xi(x_0,N, \tau) ,~~P_R(x_0) = \frac{1}{2} - \xi(x_0,N, \tau), \label{SI-44}
\end{align}
which was quoted in the Letter. Eqs.~(\ref{SI-4}-\ref{SI-44}) give the splitting probabilities for  all values of $x_0,~\tau$ and $N$ (except at the special sampling times).

\bluew{If $N$ is odd and the walker starts right at the middle, $x_0 = (N+1)/2$, then one has $\braket{E_k|x_0} = 0$ and $\sin \left( \omega_k x_0 \right)=0$ for even $k$. Consequently, we get $\rho_-(x_0) = 0$, $\widetilde{\chi}^{(-)}(z) =0$ and $\xi(x_0, N, \tau)=0$. Substituting these results into Eqs.~(\ref{SI-44}) gives $P_L(x_0)=P_R(x_0)=1/2$. This result simply reflects the symmetry of the setup: the two targets, $x_L=1$ and $x_R=N$, are equidistant from the initial position $x_0=(N+1)/2$.}

Away from this symmetry point, $\xi(x_0,N,\tau)$ can also be calculated analytically. For example, for $x_0=4$, $N=5$, Eqs.~(\ref{SI-4}-\ref{pole-eq-2}) take a form
\begin{align}
\xi(4,5, \tau) = -\frac{\sqrt{3} \sin \left(\frac{\tau}{2} \right)^2 \sin \left( \tau \right) \sin \left( \sqrt{3}  \tau \right) }{4+ \cos \left( 2  \tau \right)  + \cos \left(  \tau \right) \left[  -1+ \cos \left( \sqrt{3 } \tau \right) \left\{ -5+  \cos \left(  \tau \right) \right\} \right]}. \label{neww-eg-eq-43} 
\end{align}
Finally, we turn to the special $\tau$ values given by Eq.~\eqref{letter-res-cond}. At these special points, a separate analysis is required, involving the eigenfunctions of the unitary operator $U(t) = \exp(-iHt)$, and their associated quasi-energies \cite{SMSingh2}. For $x_0=4$, $N=5$, resonances occur at (i) $\tau = 0$, where $P_L(4) = P_R(4) = 0$; (ii) $\tau = m\pi$ ($m = 1,2,3,\ldots$), where $P_L(4) = P_R(4) = 1/4$; and (iii) $\tau = m\pi\sqrt{3}$, again with $P_L(4) = P_R(4) = 1/4$. Notice that all of them have $P_L(4)+P_R(4) < 1$. As explained before, this occurs
because a part of the system, for these special sampling times, enters into the darkspace that is decoupled from the two detectors at the measurement instances, thereby reducing the overall detection probability. These results have been shown in blue in Fig.~\ref{letter-fig-resonance} in the Letter.

{\color{black}
\begin{center}
\textit{Classical random walk under stroboscopic monitoring}
\end{center}
In the Letter, we compared our quantum results with the classical case in Eq.~\eqref{lett-unitary-eqn-classical}. It is also interesting to compare them with a classical random walk under stroboscopic monitoring, similar to the monitored quantum dynamics. In the SM \cite{SMSingh2}, we have examined the splitting probability of a monitored continuous-time, discrete-space classical random walker moving on a one dimensional chain of size $N$. 

Starting from $x_0$, the classical walker can, at any time instance, jump to one of its neighboring sites with some rate $\gamma$ which can be chosen to be unity. We further monitor the endpoints of the chain, $x_L=1$ and $x_R=N$, and check whether the walker is
at either of these sites at the stroboscopic times $\tau,~2 \tau,~3 \tau, \ldots$. This means that if the walker is located at $x_L$ or $x_R$ at the measurement times, it is absorbed and the process is stopped. Otherwise, it continues to evolve until a successful detection is made. Repeating this procedure over many realizations, we then compute the splitting probability.

When $\tau \to 0$, this setup reduces to the gambler's ruin problem in Eq.~\eqref{lett-unitary-eqn-classical}. In \cite{SMSingh2}, we show that for non-zero $\tau$, the splitting probabilities are
\begin{align}
& P_L^{(\rm cl)}(x_0) = \frac{1}{2}+ \frac{2 \sum _{k=1, \rm odd}^{N-1} ~ \cos \left( \frac{\pi k}{2N} \right)~\cos \left( \frac{\pi k}{N} \left( x_0-\frac{1}{2} \right) \right) \left( \frac{e^{\mathcal{E} _k \tau }}{1-e^{\mathcal{E} _k \tau }} \right) }{N + 4 \sum _{k=1, \rm odd}^{N-1} ~ \cos \left( \frac{\pi k}{2N} \right)^2 \left( \frac{e^{\mathcal{E} _k \tau }}{1-e^{\mathcal{E} _k \tau }} \right)},  \label{end-new-classical-pss-eq-13}   \\
& P_R^{(\rm cl)}(x_0) = \frac{1}{2}- \frac{2 \sum _{k=1, \rm odd}^{N-1} ~ \cos \left( \frac{\pi k}{2N} \right)~\cos \left( \frac{\pi k}{N} \left( x_0-\frac{1}{2} \right) \right) \left( \frac{e^{\mathcal{E} _k \tau }}{1-e^{\mathcal{E} _k \tau }} \right) }{N + 4 \sum _{k=1, \rm odd}^{N-1} ~ \cos \left( \frac{\pi k}{2N} \right)^2 \left( \frac{e^{\mathcal{E} _k \tau }}{1-e^{\mathcal{E} _k \tau }} \right)}, \label{end-new-classical-pss-eq-14} 
\end{align}
where $\mathcal{E} _  k = - 4  \sin ^2 \left( \frac{\pi k}{2 N} \right)$. For all values of $\tau$, we see that $P_L^{(\rm cl)}(x_0) + P_R^{(\rm cl)}(x_0)=1$. This is in sharp contrast to the quantum case, where at special sampling times we find $P_L(x_0)+P_R(x_0)<1$. Furthermore, we show in \cite{SMSingh2} that the splitting probabilities for non-zero $\tau$ exhibit features qualitatively similar to Eq.~\eqref{lett-unitary-eqn-classical}. For example, for any $\tau$ the proximity effect continues to hold true, namely if the walker initially is closer to the left boundary, one gets $P_L^{(\mathrm{cl})}(x_0) \geq P_R^{(\mathrm{cl})}(x_0)$ and vice versa. Moreover, the splitting probabilities in Eqs.~(\ref{end-new-classical-pss-eq-13}-\ref{end-new-classical-pss-eq-14}) are smooth monotonic functions of $\tau$ and exhibit no phase-transition-like behavior \cite{SMSingh2}. All these behaviors are in stark contrast to the monitored quantum case reported in the Letter.}

\end{widetext}

\bibliography{Bib_QRW}

\newpage
\begin{widetext}

\renewcommand{\theequation}{SM\arabic{equation}}
\setcounter{equation}{0}
\section*{SUPPLEMENTARY MATERIAL: Transition in Splitting Probabilities of Quantum Walks}
\begin{center}
Prashant Singh, David A. Kessler and Eli Barkai\\
\textit{Department of Physics, Bar-Ilan University, Ramat Gan 52900, Israel}
\end{center}

The Supplementary Material contains the following: 
\begin{enumerate}
\item Derivation of the first-detection amplitudes in Sec.~\ref{appen-sec-FDT},

\item Proof of the mapping to single-target detection in Sec.~\ref{appen-SM-map},

\item Derivation of the resonance conditions in Sec.~\ref{appen-SM-DS},

\item Derivation of the eigenvalues $\{ \lambda_{\pm} \}$ of the survival operators $\mathcal{S}_{\pm}=\Big( \mathbbm{1}-\ket{d_{\pm}} \bra{d_{\pm}}\Big) U(\tau) $ in Sec.~\ref{appen-SM-eigenvalue}, 

\item Expressions of the splitting probability in terms of $\{ \lambda_{\pm} \}$ in Sec.~\ref{sec-SM-relation}, 

\item Spectral origin of the transition in the splitting probability in Sec.~\ref{sec-SM-spectral},

\item System size independence of $P_L(x_0)$ and $P_R(x_0)$ for large system sizes in Sec.~\ref{appen-largeN},

\item Classical random walk with stroboscopic monitoring in Sec.~\ref{appen-classical}.
\end{enumerate}

\section{Derivation of the first-detection amplitudes}
\label{appen-sec-FDT}
{\color{black}
In Eq.~(3) of the Letter, we showed that the first-detection amplitudes $\phi_n^{L}$ and $\phi_n^{R}$ are expressed in terms of the survival operator $\mathcal{S} = \Big[ \Big(\mathbbm{1}-\ket{x_L}  \bra{x_L} - \ket{x_R}  \bra{x_R} \Big) U(\tau) \Big]$. We will derive this relation here.

To begin with, let us consider the first measurement at $n=1$. The state of the system just before the measurement is 
\begin{align}
\ket{\psi(\tau^-) }= U(\tau)  \ket{\psi _0}. \label{unitary-eqn-2}
\end{align}
Here $\tau^-$ is the instant just before the measurement. We then do the first measurement at $\ket{x_L}$ at time $t = \tau$, and the corresponding probability of detection is
\begin{align}
Q^{(L)}_1 \left(  \psi _0 \right) = \big| \langle x_L \ket{\psi(\tau^-) } \big|^2 = \big| \bra{x_L}U(\tau) \ket{\psi _0} \big|^2.  \label{unitary-eqn-3}
\end{align}
We denote by $Q_{n}^{(\alpha)}\left(  \psi _0 \right)$, with ${\alpha} \in \{L,R \}$, the conditional probability of first-detection of the $\ket{x_{\alpha}}$ target state at the $n$-th measurement, with no detection occurring in any of the two target states at the earlier attempts.


Turning to Eq.~\eqref{unitary-eqn-3}, if the system is detected at $\ket{x_L}$, the process terminates. If it is not detected, then according to the standard projection postulate of quantum mechanics \cite{Cohenbook}, the state vector collapses onto the subspace orthogonal to $\ket{x_L}$, i.e., the component along $\ket{x_L}$ is projected out. Therefore, right after a null measurement of $\ket{x_L}$, the state vector at time $t = \tau^+$ collapses to \cite{Cohenbook}
\begin{align}
\ket{\psi(\tau^+)}
= \mathcal{N}\Big(\mathbbm{1}-\ket{x_L}\bra{x_L}\Big)\ket{\psi(\tau^-)}
= \mathcal{N}\Big(\mathbbm{1}-\ket{x_L}\bra{x_L}\Big)U(\tau)\ket{\psi_0}.
\label{unitary-eqn-4}
\end{align}
where $\mathcal{N} = 1 / \sqrt{1-Q^{(L)}_1\left(  \psi _0 \right)} $ gives the correct normalisation $\langle \psi(\tau^+ )\ket{ \psi(\tau ^+)} = 1$. Thus the state after measurement is correctly normalized. 

Given the normalized state $\ket{ \psi(\tau^+ )} $, we next perform an immediate projective measurement of the second target state $\ket{x_R}$. The probability of detection, conditioned that the first measurement of $\ket{x_L}$ was a failure, is
\begin{align}
Q^{(R)}_1\left(  \psi _0 \right) = \big| \langle x_R \ket{\psi(\tau^+) } \big|^2 = \frac{ \big| \bra{x_R} \Big( \mathbbm{1} - \ket{x_L} \bra{x_L}\Big)   U(\tau) \ket{\psi _0} \big|^2}{\left( 1-Q^{(L)}_1  \left(  \psi _0 \right) \right)  }.  \label{unitary-eqn-5}
\end{align} 
Using the orthogonality condition $\langle x_R \ket{x_L} = 0$, we obtain
\begin{align}
Q^{(R)}_1 \left(  \psi _0 \right) = \frac{ \big| \bra{x_R}  U(\tau) \ket{\psi _0} \big|^2}{\left(  1-Q^{(L)}_1 \left(  \psi _0 \right) \right) }.  \label{unitary-eqn-6}
\end{align}
This represents the conditional probability that the system is detected at the $\ket{x_R}$ state at $t=\tau$, given that it was not detected at $\ket{x_L}$ in the earlier measurement. However, if the detection at $\ket{x_R}$ is also not successful, then the state of the system at the end of this measurement becomes
\begin{align}
\ket{\psi(\tau^{++})} = \frac{\Big(\mathbbm{1} - \ket{x_R} \bra{x_R} \Big) \ket{\psi \left( \tau^+\right)  }}{\sqrt{1-Q^{(R)}_1 \left(  \psi _0 \right)   }} = \frac{\Big(\mathbbm{1} - \ket{x_L} \bra{x_L}-\ket{x_R} \bra{x_R} \Big) U(\tau) \ket{\psi_0 }}{\sqrt{\left( 1-Q^{(L)}_1  \left(  \psi _0 \right) \right)  \left( 1-Q^{(R)}_1  \left( \psi _0 \right) \right)} }, \label{unitary-eqn-7}
\end{align}
where $`++$' in $\ket{\psi(\tau^{++})}$ denotes that two measurements were performed at time $\tau$. This state is again correctly normalized. Repeating the same procedure, the state of the system just before the \(n\)-th measurement, conditioned on no detection of either target state in the previous $(n-1)$ measurements, is
\begin{align}
    \ket{\psi(n \tau^-)} = \frac{ U(\tau) \Big[ \Big(\mathbbm{1}- \ket{x_L} \bra{x_L}-\ket{x_R} \bra{x_R} \Big) U(\tau) \Big]^{n-1}\ket{ \psi _0}  }{\prod_{j=1}^{n-1}\sqrt{\Big[ \left( 1-Q^{(L)}_{j}  \left(  \psi _0 \right) \right)  \left( 1-Q^{(R)}_{j} \left(  \psi _0 \right) \right) \Big]}}.
\end{align}
Here again the state is correctly normalized, $\braket{\psi(n \tau^-)|\psi(n \tau^-)}=1$. Then the conditional probability $Q_{n}^{(\alpha)}(\psi _0)$ of first detecting the $\ket{x_{\alpha}}$ state at the $n$-th measurement attempt can be shown to be
\begin{align}
& Q^{(L)}_n \left(  \psi _0 \right) =  \frac{\big| \bra{x_L} U(\tau) \Big[ \Big(\mathbbm{1}-\ket{x_L} \bra{x_L}-\ket{x_R} \bra{x_R} \Big) U(\tau) \Big]^{n-1}\ket{ \psi _0}   \big|^2}{\prod_{j=1}^{n-1}\Big[ \left( 1-Q^{(L)}_{j}  \left(  \psi _0 \right) \right)  \left( 1-Q^{(R)}_{j} \left(  \psi _0 \right) \right) \Big] }, \label{unitary-eqn-11} \\
& Q^{(R)}_n\left( \psi _0 \right)  =  \frac{\big| \bra{x_R} U(\tau) \Big[ \Big(\mathbbm{1}- \ket{x_L} \bra{x_L}-\ket{x_R} \bra{x_R} \Big) U(\tau) \Big]^{n-1}\ket{ \psi _0}   \big|^2}{\left( 1-Q^{(L)}_{n} (\psi_0)\right)~\prod_{j=1}^{n-1}\Big[  \left( 1-Q^{(L)}_{j} \left(  \psi _0 \right) \right)  \left( 1-Q^{(R)}_{j}  \left(  \psi _0 \right) \right) \Big]}. \label{unitary-eqn-12} 
\end{align} 
For the second equation, we get an additional factor of $\left( 1-Q^{(L)}_{n} (\psi_0)\right)$ in the denominator because $Q^{(R)}_n\left( \psi _0 \right)$ requires a failed detection at $\ket{x_L}$ at $t=n \tau$.

Remember that both $ Q^{(L)}_n \left(  \psi _0 \right) $ and $ Q^{(R)}_n \left(  \psi _0 \right) $ denote conditional detection probabilities, defined under the requirement that no successful detection has occurred in any of the preceding measurement attempts. From these, the first-detection probabilities of the two target states at the $n$-th stroboscopic attempt are
 \begin{align}
\resizebox{0.8\textwidth}{!}{$
\begin{aligned}
& F_{n}^{(L)}  \left(  \psi _0 \right) =   Q^{(L)}_n \left(  \psi _0 \right) ~\prod_{j=1}^{n-1}\Big[ \left( 1-Q^{(L)}_{j} \left(  \psi _0 \right)  \right)  \left( 1-Q^{(R)}_{j}  \left( \psi _0 \right)\right) \Big], \\[6pt]
& F_{n}^{(R)}  \left(  \psi _0 \right) =   Q^{(L)}_n  \left(  \psi _0 \right) ~\left( 1-Q^{(L)}_{n} \left( \psi _0 \right) \right)~\prod_{j=1}^{n-1}\Big[  \left( 1-Q^{(L)}_{j}  \left(  \psi _0 \right) \right)  \left( 1-Q^{(R)}_{j}  \left( \psi _0 \right) \right) \Big].
\end{aligned}
$} \label{unitary-eqn-13hau18}
\end{align}
In the first relation, the product term 
$\prod_{j=1}^{n-1} \big[\cdots\big]$ represents the probability that all previous $(n-1)$ measurement attempts were unsuccessful, 
meaning that the system was not detected at $\ket{x_L}$ or $\ket{x_R}$. This factor is required when converting from the conditional probability $Q_n^{(\alpha)}(\psi _0)$, which assumes no earlier detections, to the 
unconditional probability $F_n^{(\alpha)}(\psi _0)$, which explicitly accounts for the likelihood of this condition being satisfied. In the first relation, this product alone is sufficient, as we only need the system to be detected at the $\ket{x_L}$ state at the $n$-th attempt. In contrast, there is an additional factor of $(1 - Q_n^{(L)} (\psi _0))$ in the second relation. This factor accounts for the failed detection of $\ket{x_L}$ at the $n$-th attempt, so that the system gets detected at $\ket{x_R}$.

Plugging Eqs.~(\ref{unitary-eqn-11}-\ref{unitary-eqn-12}) in Eq.~\eqref{unitary-eqn-13hau18} yields
\begin{align}
& F_{n}^{(\alpha)}\left(  \psi _0 \right) = \big|  \phi_n^{(\alpha)}   \big|^2,~~\text{with }   \alpha \in \{L,R \}. \label{unitary-eqn-13} , \\
& \phi_n^{(\alpha} = \bra{x_{\alpha}} U(\tau) \Big[ \Big(\mathbbm{1}-\ket{x_L} \bra{x_L}-\ket{x_R} \bra{x_R} \Big) U(\tau) \Big]^{n-1}\ket{\psi _0}. \label{unitary-eqn-13-sknqi}
\end{align}
This result has been quoted in the Letter in Eq.~(3).}

\section{Derivation of the mapping to single-target detection}
\label{appen-SM-map}
\indent
In the Letter, we pointed out that the first-detection amplitude $\phi_n^{(\alpha)}$ for detection at the boundary $x_{\alpha}$ $(\alpha \in \{L,R\})$, given the absence of detection at the opposite boundary $x_{\overline{\alpha}}$ (with $\overline{\alpha}=R$ for $\alpha=L$ and vice versa), satisfies a quantum mapping for the parity-symmetric Hamiltonians. This mapping connects the dual-target detection amplitudes to two distinct auxiliary single-target first-detection amplitudes $\chi^{(\pm)}_n$ as follows
\begin{align}
\phi _{n}^{\left(  L \right)}  = \frac{\left( \chi _{n}^{\left(+\right)}  + \chi _{n}^{(-)} \right)}{\sqrt{2}}  ,~~\phi _{n}^{\left(  R \right)}  = \frac{\left( \chi _{n}^{\left(+\right)}  - \chi _{n}^{(-)} \right)}{\sqrt{2}}, \label{supp-SR-eq-1}
\end{align}
where amplitudes $\chi^{(\pm)}_n$ are given by
\begin{align}
\chi^{(\pm)}_n = \bra{d_{\pm}} U(\tau) \Big[ \Big( \mathbbm{1}-\ket{d_{\pm}} \bra{d_{\pm}}\Big) U(\tau) \Big]^{n-1}\ket{ \psi _0}. \label{supp-SR-eq-2}
\end{align}
Here $U(\tau)=\exp(-iH\tau)$, $\ket{d_{\pm}} = \left( \ket{x_L} \pm \ket{x_R}  \right)/ \sqrt{2}$, $\ket{x_L}=1$ and $\ket{x_R}=N$. Before deriving Eq.~\eqref{supp-SR-eq-1}, we recall some basic properties of the parity (reflection) operator $\mathcal{P}$, defined by its action on the position basis as
\begin{equation}
\mathcal{P}\ket{x} = \ket{N+1-x},  \label{supp-SR-eqn-1}
\end{equation}
where recall $x=1,2, \ldots,N$. The parity operator is its own inverse, $\mathcal{P}^{-1}=\mathcal{P}$, and satisfies $\mathcal{P}^2=\mathbbm{1}$. For Hamiltonians possessing the reflection symmetry, the parity operator commutes with the Hamiltonian
\begin{equation}
[H,\mathcal{P}] = 0,~~~\implies \mathcal{P}H\mathcal{P} = H.
\end{equation}
As a consequence, the unitary operator also satisfies
\begin{equation}
\mathcal{P}U(\tau)\mathcal{P} = U(\tau). \label{supp-SR-eqn-2}
\end{equation}
We also have $\mathcal{P} \ket{d_{\pm}} = \pm \ket{d_{\pm}}$. Combining this relation with Eq.~\eqref{supp-SR-eqn-2} gives
\begin{align}
\braket{d_+|U(\tau)|d_-} = \braket{d_-|U(\tau)|d_+} = 0, \label{supp-SR-eqn-3}
\end{align}
for the parity-symmetric Hamiltonians. With these results at disposal, we proceed to derive the mapping in Eq.~\eqref{supp-SR-eq-1}. 

Let us analyze the expression of $\phi_n^{(\alpha)}$ given in Eq.~\eqref{unitary-eqn-13-sknqi}
\begin{equation}
\phi _n^{(\alpha)} =  \bra{x_{\alpha}} U(\tau)  \Big[ \Big( \mathbbm{1}-\ket{x_L} \bra{x_L}- \ket{x_R} \bra{x_R} \Big) U(\tau) \Big]^{n-1}\ket{ \psi _0},~~\alpha \in \{L,R \}. \label{supp-SR-eq-3}
\end{equation}
For $n=1$, $\phi _n^{(L)}$ and $\phi _n^{(R)}$ readily satisfy Eq.~\eqref{supp-SR-eq-1}. On the other hand, for $n=2$, one gets
\begin{align}
&\phi _2^{(L)} = \frac{1}{\sqrt{2}} \left[ \chi_2^{(+)} + \chi_2^{(-)} - \braket{d_+|U(\tau)|d_-} \braket{d_-|U(\tau)|\psi_0} -\braket{d_-|U(\tau)|d_+} \braket{d_+|U(\tau)|\psi_0} \right],  \label{supp-SR-eq-4} \\
& \phi _2^{(R)} = \frac{1}{\sqrt{2}} \left[ \chi_2^{(+)} - \chi_2^{(-)} - \braket{d_+|U(\tau)|d_-} \braket{d_-|U(\tau)|\psi_0} +\braket{d_-|U(\tau)|d_+} \braket{d_+|U(\tau)|\psi_0} \right]. \label{supp-SR-eq-5}
\end{align}
Substituting Eq.~\eqref{supp-SR-eqn-3}, we see that Eqs.~\eqref{supp-SR-eq-4} and \eqref{supp-SR-eq-5} reduce to our mapping in Eq.~\eqref{supp-SR-eq-1}. Hence, the mapping is satisfied for $n=2$. 

We now use induction to prove that Eq.~\eqref{supp-SR-eq-1} holds for general $n$. Assume that the mapping is valid for $n=m$. Then
\begin{align}
\bra{x_{L}} U(\tau)  \Big[ \Big( \mathbbm{1}-\ket{x_L} \bra{x_L}- \ket{x_R} \bra{x_R} \Big) U(\tau) \Big]^{m-1} & =  \frac{1}{\sqrt{2}} \left[   \bra{d_+} U(\tau) \Big[ \Big( \mathbbm{1}-\ket{d_{+}} \bra{d_{+}}\Big) U(\tau) \Big]^{m-1} \right.  \nonumber \\
& \left. +    \bra{d_-} U(\tau) \Big[ \Big( \mathbbm{1}-\ket{d_{-}} \bra{d_{-}}\Big) U(\tau) \Big]^{m-1}  \right],
\end{align}
where we have focused on the left boundary for simplicity. The derivation for the right one proceeds analogously. We plug this relation in Eq.~\eqref{supp-SR-eq-3} for $n=(m+1)$ to yield
\begin{align}
\phi _{m+1}^{(L)} = & \frac{1}{\sqrt{2}} \left[ \chi_{m+1}^{(+)} + \chi_{m+1}^{(-)} - \braket{d_+|U(\tau) \Big[ \left( 1-\ket{d_+}\bra{d_+}\right)U(\tau)\Big]^{m-1}   |d_-} \braket{d_-|U(\tau)|\psi_0} \right.\nonumber \\
&~~~~~\left.-\braket{d_-|U(\tau) \Big[ \left( 1-\ket{d_-}\bra{d_-}\right)U(\tau)\Big]^{m-1}  |d_+} \braket{d_+|U(\tau)|\psi_0} \right]. \label{supp-SR-eq-6}
\end{align}
Since the projector $\ket{d_\pm}\bra{d_\pm}$ is parity invariant, the operator
$U(\tau)[(\mathbbm{1}-\ket{d_\pm}\bra{d_\pm})U(\tau)]^{m-1}$ is parity symmetric, i.e.,
\begin{align}
\mathcal{P} U(\tau)[(\mathbbm{1}-\ket{d_\pm}\bra{d_\pm})U(\tau)]^{m-1} \mathcal{P} = U(\tau)[(\mathbbm{1}-\ket{d_\pm}\bra{d_\pm})U(\tau)]^{m-1}.
\end{align}
It then follows that
\begin{align}
& \bra{d_+}
U(\tau)\big[(\mathbbm{1}-\ket{d_\pm}\bra{d_\pm})U(\tau)\big]^{m-1}
\ket{d_-}
= 0 , \\
& \bra{d_-}
U(\tau)\big[(\mathbbm{1}-\ket{d_\pm}\bra{d_\pm})U(\tau)\big]^{m-1}
\ket{d_+}
= 0.
\end{align}
Plugging them in Eq.~\eqref{supp-SR-eq-6}, we finally obtain
\begin{align}
\phi _{m+1}^{(L)} = & \frac{1}{\sqrt{2}} \left[ \chi_{m+1}^{(+)} + \chi_{m+1}^{(-)} \right].
\end{align}
This completes the induction and establishes the mapping~\eqref{supp-SR-eq-1} for arbitrary $n$. The derivation for the right boundary also follows in the same way.

\section{Resonance conditions}
\label{appen-SM-DS}
{\color{black}
In the Letter, we stated that the condition for resonance at special sampling times is phase matching within the same parity sector. To derive this condition, we begin by summing the square moduli of $\phi_n^{(L)}$ and $\phi_n^{(R)}$ in Eq.~\eqref{supp-SR-eq-1} over $n$ to yield the splitting probabilities
\begin{align}
& P_L(x_0) = \sum _{n=1}^{\infty} \big| \phi_n^{(L)} \big|^2 = \frac{1}{2} \left[ \rho _+(x_0) + \rho _-(x_0) \right]+  ~\text{Re} \left[ \sum_{n=1}^{\infty} \chi _{n}^{\left( + \right)} ~\left( \chi _{n}^{\left( - \right)} \right)^* \right], \label{SM-SI-1} \\
& P_R(x_0) = \sum _{n=1}^{\infty} \big| \phi_n^{(R)} \big|^2 = \frac{1}{2} \left[ \rho _+(x_0) + \rho _-(x_0) \right]- ~\text{Re} \left[ \sum_{n=1}^{\infty} \chi _{n}^{\left( + \right)} ~\left( \chi _{n}^{\left( - \right)} \right)^* \right]. \label{SM-SI-2}
\end{align}
Here, $\rho _{\pm}(x_0) = \sum _{n=1}^{\infty} \big| \chi _{n}^{\left( \pm \right)}  \big|^2  $ are the total detection probabilities in the auxiliary single-target problems. To compute them, we leverage methods developed in earlier works \cite{QW6, QW12,QW16, QW21, QW7, QW8,QW13}. These methods use the eigenstates of the unitary operator $U(\tau)=\exp(-iH \tau )$. The eigenvalues of $U(\tau)$ take the form 
$e^{-i\lambda_k}$, where 
\begin{align}
\lambda_k = E_k \tau \text{ mod }2 \pi, \label{neww-eg-eq-1q}
\end{align}
are the quasienergies with $k=1,2,...,N$. Recall that, for the nearest-neighbor tight-binding Hamiltonian studied in the Letter, we have
\begin{align}
E_{k}=-2  \cos \left( \omega_k \right),~~\ket{E_k} = \sqrt{\frac{2}{N+1}}~ \sum _{x=1}^{N} ~\sin \left( \omega _k x \right)~\ket{x}, \label{SSPM-eq-ev-2}
\end{align}
where $\omega_k=\frac{\pi k}{N+1} $. All energy levels are nondegenerate. 

For most choices of $\tau$, there is a one-to-one correspondence between the energy eigenvalues $\{E_k\}$ of $H$ and the quasienergies $\{ \lambda_k \}$. At certain special values of $\tau$, however, different energies $\{E_k, E_{\ell} \}$ can yield the same phase factor $e^{-iE_k\tau} = e^{-iE_{\ell}\tau}$, leading to the same quasienergy value. This gives rise to degeneracies in the quasienergy spectrum, and hence fewer distinct values of $\lambda_k$ compared to the number of energy eigenvalues. Let $\left\{ \ket{E_{k,m}} \right\}$, with $m=1,2,\ldots,g_k$, be the degenerate states associated with the same $\lambda _k$ and $g_k \ge 1$ being the degeneracy factor.

In the quasienergy basis, $\rho _{\pm} \left( x _0 \right) $ take the form \cite{QW16}
\begin{align}
& \rho _{+}(x _0) = \sum_{ \{ \ket{E_{k,m} } \} \in ~\rm even ~state } 
      \frac{\big| \sum_{m=1}^{g_k} 
         \braket{d_{+} \big| E_{k,m}} \braket{E_{k,m} \big| x _0} \big|^2}
           { \sum_{m=1}^{g_k} \big| \braket{d_{+} \big| E_{k,m}} \big|^2}, \label{SM-new-pseb-eq-2} \\
& \rho _{-}(x _0) = \sum_{ \{ \ket{E_{k,m}}\} \in ~\rm odd~state } 
      \frac{\big| \sum_{m=1}^{g_k} 
         \braket{d_{-} \big| E_{k,m}} \braket{E_{k,m} \big| x _0} \big|^2}
           { \sum_{m=1}^{g_k} \big| \braket{d_{-} \big| E_{k,m}} \big|^2}. \label{SM-new-pseb-eq-3}           
\end{align}
For $\rho _{+}(x _0) $, we have used the notation ``$\{ \ket{E_{k,m} } \} \in ~\rm even ~state $" in the summation to indicate that the summation is restricted only to the even-parity quasi-energy states, since only these states will have a non-vanishing overlap with $\ket{d_{+}}$ (one has $\braket{d_+|E_{k,m}}=0$ if $\ket{E_{k,m}}$ has odd parity). Similarly, for $\rho _{-}(x _0) $, only odd-parity quasi-energy states contribute and hence the notation ``$\{ \ket{E_{k,m} } \} \in ~\rm odd ~state $". 

Let us next turn to the mixed term in Eqs.~(\ref{SM-SI-1}-\ref{SM-SI-2}). One can express it as
\begin{align}
\xi(x_0, N, \tau) & = \sum_{n=1}^{\infty} \text{Re}  \left[ \chi^{(+)}(n) (\chi^{(-)}(n))^* \right] =  \oint_{|z|=1} \frac{d z}{2  \pi i z}~ \text{Re}\left[ \widetilde{\chi}^{(+)}(z)  \left( \widetilde{\chi}^{(-)}(1/z) \right)^*  \right], \label{SP-SR-eq-6}
\end{align}
where $\widetilde{\chi}^{(\pm)}(z) = \sum_{n=1}^{\infty} z^n \chi^{(\pm)}(n)$ are the auxiliary generating functions. Since for  any two complex numbers $z_1$ and $z_2$, we have $\text{Re}[z_1 z_2^*] = \text{Re}[z_1^* z_2]$, Eq.~\eqref{SP-SR-eq-6} can also be written as
\begin{align}
\xi(x_0, N, \tau) & = \sum_{n=1}^{\infty} \text{Re}  \left[ (\chi^{(+)}(n))^* \chi^{(-)}(n) \right] =  \oint_{|z|=1} \frac{d z}{2  \pi i z}~ \text{Re}\left[\left( \widetilde{\chi}^{(+)}(1/z) \right)^*   \widetilde{\chi}^{(-)}(z)   \right]. \label{SP-SR-eq-6-nn}
\end{align}
As shown in Sec.~\ref{sec-SM-relation}, these two forms lead to distinct but equivalent spectral representations of $\xi(x_0,N,\tau)$.

The generating functions, $\widetilde{\chi}^{(\pm)}(z)$, can be written in terms of the quasienergy spectrum as follows
\begin{align}
& \widetilde{\chi}^{(+)}(z) =   \frac{ \sum_{\{ \ket{E_{k,m}} \} \in ~ \rm{even~state}} \sum_{m=1}^{g_k}   \braket{d_{+} |E_{k,m} \rangle \langle E_{k,m}|x _0}  \frac{z e^{-i \lambda_k}}{1-z e^{-i \lambda_k}}}{ \sum_{\{ \ket{E_{k,m}} \} \in ~ \rm{even~state}} \sum _{m=1}^{g_k} \big| \braket{d_+|E_{k,m}} \big|^2 \frac{1}{1-z e^{-i \lambda_k}}}, \label{SP-SR-eq-7} \\
& \widetilde{\chi}^{(-)}(z) =   \frac{ \sum_{\{ \ket{E_{k,m}} \} \in ~ \rm{odd~state}} \sum_{m=1}^{g_k}   \braket{d_{-} |E_{k,m} \rangle \langle E_{k,m}|x _0}  \frac{z e^{-i \lambda_k}}{1-z e^{-i \lambda_k}}}{ \sum_{\{ \ket{E_{k,m}} \} \in ~ \rm{odd~state}} \sum _{m=1}^{g_k} \big| \braket{d_-|E_{k,m}} \big|^2 \frac{1}{1-z e^{-i \lambda_k}}}. \label{SP-SR-eq-8} 
\end{align}	
Again, we see that $\widetilde{\chi}^{(+)}(z)$ contains only even-parity states, whereas $\widetilde{\chi}^{(-)}(z)$ involves only odd-parity states. Plugging these results in Eq.~\eqref{SP-SR-eq-6} or Eq.~\eqref{SP-SR-eq-6-nn} yield the mixed term $\xi(x_0,N,\tau)$, and combining it with $\rho_{\pm}(x_0)$ in Eqs.~(\ref{SM-new-pseb-eq-2},~\ref{SM-new-pseb-eq-3}), we obtain the exact splitting probability for all sampling times. 

As mentioned before, for generic (non-special) values of $\tau$, there is a one-to-one correspondence between the energy eigenvalues and the quasienergies. Since all energy levels are nondegenerate [see Eq.~\eqref{SSPM-eq-ev-2}], the degeneracy factor in the quasienergy spectrum is unity, \textit{i.e.,} $g_k=1$. Moreover, Eq.~\eqref{SSPM-eq-ev-2} shows that $\ket{E_k}$ has even parity for odd $k$ and odd parity for even $k$, where recall $k \in \{1,2,\ldots,N \}$. Then the expressions of $\rho_{\pm}(x_0)$ and $\widetilde{\chi}^{(\pm)}(z) $ reduce to 
\begin{align}
\rho _+(x _0) = \sum_{\substack{k=1 \\ k \in  \,\text{odd}}}^{N} \big| \braket{E_k | x _0} \big|^2,~&~\rho _-(x _0) = \sum_{\substack{k=1 \\ k \in  \,\text{even}}}^{N} \big| \braket{E_k | x _0} \big|^2, \label{aux-eq-4ji} \\
\widetilde{\chi}^{(+)}(z) =\frac{\sum _{k \in ~\rm{odd}} \sin (\omega _k) \sin (\omega_k x_0 ) \left( \frac{z e^{2 i  \tau \cos (\omega _k)}}{1-z ~e^{2 i \tau \cos (\omega _k)}}  \right)  }{  \sqrt{2} \sum _{k \in ~\rm{odd}} \sin ^2 (\omega _k)  \left( \frac{1}{1-z ~e^{2 i \tau \cos (\omega _k)}}  \right)      },~&~\widetilde{\chi}^{(-)}(z) =\frac{\sum _{k \in ~\rm{even}} \sin (\omega _k) \sin (\omega_k x_0 ) \left( \frac{z e^{2 i  \tau \cos (\omega _k)}}{1-z ~e^{2 i \tau \cos (\omega _k)}}  \right)  }{ \sqrt{2}  \sum _{k \in ~\rm{even}} \sin ^2 (\omega _k)  \left( \frac{1}{1-z ~e^{2 i  \tau \cos (\omega _k)}}  \right) }. \label{SI-3j81}
\end{align}
Substituting the generating functions in Eq.~\eqref{SP-SR-eq-6} or Eq.~\eqref{SP-SR-eq-6-nn} gives $\xi(x_0, N, \tau)$ and this result along with Eq.~\eqref{aux-eq-4ji} specify the splitting probabilities in Eqs.~(\ref{SM-SI-1}-\ref{SM-SI-2})
\begin{align}
P_L(x_0) = \frac{1}{2} + \xi(x_0,N, \tau) ,~~P_R(x_0) = \frac{1}{2} - \xi(x_0,N, \tau). \label{SioanxI-44}
\end{align}
Notice that, if $N$ is odd and the walker starts right at the middle, $x_0 = (N+1)/2$, then one has $\braket{E_k|x_0} = 0$ and $\sin \left( \omega_k x_0 \right)=0$ for even $k$. Consequently, we get $\widetilde{\chi}^{(-)}(z) =0$ and $\xi(x_0, N, \tau)=0$. Substituting these results into Eq.~(\ref{SioanxI-44}) gives $P_L(x_0)=P_R(x_0)=1/2$. This result simply reflects the symmetry of the setup: the two targets, $x_L=1$ and $x_R=N$, are equidistant from the initial position $x_0=(N+1)/2$.

Eq.~\eqref{SioanxI-44} is valid for generic $\tau$ except for the special cases. Then a separate analysis is required. As explained before, for these special values, two or more quasi-energies from the same parity-sector can become equal
\begin{align}
\lambda _k = \lambda_{\ell} \implies \left( E_k-E_{\ell} \right) \tau = 0~\text{mod } 2 \pi \label{resonance-eqn}
\end{align}
where $k \neq \ell$ and $\ket{E_k}$ and $\ket{E_{\ell}}$ have the same parity. Such coincidences reduce the number of non-degenerate quasi-energies and one has $g_k > 1$.
At such special values of $\tau$, the structure of the summations appearing in the expressions of the auxiliary quantities, $\rho _{\pm}(x_0)$ and $\widetilde{\chi} ^{(\pm)}(z)$ in Eqs.~(\ref{SM-new-pseb-eq-2}-\ref{SM-new-pseb-eq-3}-\ref{SP-SR-eq-7}-\ref{SP-SR-eq-8}), changes, as multiple terms coalesce into a single contribution. This coalescence leads to a sudden jump in the values of these quantities, corresponding to sharp transitions in the splitting probabilities $P_L(x_0)$ and $P_R(x_0)$. Coincidences of quasienergies between different parity-sectors, however, leave the summation structure unchanged, and thus do not produce resonances. Eq.~\eqref{resonance-eqn} is thus the resonance condition which is quoted in the Letter.}

\section{Eigenvalues of the single-target survival operators}
\label{appen-SM-eigenvalue}
In Eq.~\eqref{supp-SR-eq-6}, we saw that the detection amplitudes $\chi_{n}^{(\pm)}$ are characterized by the survival operators 
\begin{align}
\mathcal{S}_{\pm}=\Big( \mathbbm{1}-\ket{d_{\pm}} \bra{d_{\pm}}\Big) U(\tau). \label{SM-eq-ev-1}
\end{align}
The eigenspectrum of $\mathcal{S}_{\pm}$ determines the behavior of $\chi_{n}^{(\pm)}$. Here, we examine their eigenvalues for the nearest-neighbor tight-binding Hamiltonian for non-resonant sampling times, i.e., Eq.~\eqref{resonance-eqn} is not satisfied. These results are summarized in Table~\ref{tab:Spm-eigs}, and we derive them below. For simplicity, we present the derivation for $\mathcal{S}_{-}$. The same steps can be carried out for $\mathcal{S}_{+}$.

The eigenvalues and the eigenstates of the Hamiltonian are given by
 \begin{align}
E_{k}=-2  \cos \left( \omega_k \right),~~\ket{E_k} = \sqrt{\frac{2}{N+1}}~ \sum _{x=1}^{N} ~\sin \left( \omega_k x \right)~\ket{x}, \label{SM-eq-ev-2}
\end{align}
where $\omega_k= \pi k / (N+1)$ with $k = 1,2, \ldots,N$. The target state $\ket{d_{-}} = (\ket{1}-\ket{N})/\sqrt{2}$ becomes
\begin{align}
\ket{d_{-}} =  \frac{2}{\sqrt{N+1}} \sum_{\substack{k=1 \\ k \in  \,\text{even}}}^{N} \sin \left( \frac{\pi k}{N+1} \right)~\ket{E_k}. \label{SM-eq-ev-3}
\end{align}
Let $\lambda_{-}$ denote the eigenvalue of $\mathcal{S}_-$. It satisfies the characteristic equation
\begin{align}
& \text{det}\left( \lambda_{-} \mathbbm{1} -\mathcal{S}_-  \right)=0, \\
\implies & \text{det} \left(\lambda_{-} \mathbbm{1}-U(\tau)+\ket{d_-}\bra{d_-} U(\tau)\right) = 0, \\
\implies & \text{det} \left( \lambda_- \mathbbm{1}-U(\tau) \right) \times \left( 1 + \left\langle d_- \bigg| \frac{U(\tau)}{\lambda_- \mathbbm{1}-U(\tau)} \bigg|d_- \right\rangle \right)=0.
\label{SM-eq-ev-4}
\end{align}
In the last line, we have used the matrix determinant lemma: For an invertible square matrix $A$ and column vectors $\ket{u}$ and $\ket{v}$ 
\begin{align}
\text{det} \left( A+\ket{u} \bra{v} \right) = \text{det}
(A) \times \left(  1 + \braket{v|A^{-1}|u}  \right). \label{SM-eq-ev-5}
\end{align}
Eq.~\eqref{SM-eq-ev-4} implies that 
\begin{align}
\text{det} \left( \lambda_- \mathbbm{1}-U(\tau) \right) = 0, \quad \text{or } \left( 1 + \left\langle d_- \bigg| \frac{U(\tau)}{\lambda_- \mathbbm{1}-U(\tau)} \bigg|d_- \right\rangle \right) = 0. \label{SM-eq-ev-6}
\end{align}
The first condition yields $\lambda_- = e^{-i E_k \tau}$, i.e., the eigenvalues of $U(\tau)$. From Eq.~\eqref{SM-eq-ev-1}, we see that such a solution occurs only if the corresponding eigenmode of $U(\tau)$ is orthogonal to the $\ket{d_-}$. Eq.~\eqref{SM-eq-ev-3} tells us that $\braket{E_k|d_-} = 0$ if $k$ is odd. This means that
\begin{align}
\lambda_- = \{ e^{-iE_k \tau}\}, \quad \text{where } k \text{ is odd}. \label{SM-eq-ev-7}
\end{align}
These account for $N/2$ eigenvalues of $\mathcal{S}_-$ when $N$ is even, and $(N+1)/2$ eigenvalues when $N$ is odd. The remaining eigenvalues follow from the second condition in Eq.~\eqref{SM-eq-ev-6}
\begin{align}
\lambda_- \left\langle d_- \bigg| \frac{ \mathbbm{1}}{\lambda_- \mathbbm{1}-U(\tau)} \bigg|d_- \right\rangle = 0. \label{SM-eq-ev-8}
\end{align}
The trivial solution is 
\begin{align}
\lambda_- = 0. \label{SM-eq-ev-99}
\end{align}
Other solutions are
\begin{align}
& \left\langle d_- \bigg| \frac{ \mathbbm{1}}{\lambda_- \mathbbm{1}-U(\tau)} \bigg|d_- \right\rangle = 0,  
\implies   \sum_{\substack{k=1 \\ k \in  \,\text{even}}}^{N} \frac{\sin^2 (\omega _k) }{\left( \lambda_-- ~e^{2 i  \tau \cos (\omega _k) } \right)}   =0, \label{SM-eq-ev-9}
\end{align}
where $\omega_k = \pi k /(N+1)$. Eq.~\eqref{SM-eq-ev-9} can be interpreted in the language of the classical charge theory \cite{QW21, QW13}. The points $e^{2 i \tau \cos(\omega_k)}$ (for even $k$) lie on the unit circle and act as sources with positive weights $\sin^2(\omega_k)$, while Eq.~\eqref{SM-eq-ev-9} is precisely the requirement that the net force field vanishes at $\lambda_-$, i.e., $\lambda_-$ is an equilibrium point of the forces generated by these charges. Solving Eq.~\eqref{SM-eq-ev-9} generates the $(N/2-1)$ eigenvalues of $\mathcal{S}_-$ for even $N$ and $(N-3)/2$ for odd $N$ located in the unit disc ($|\lambda _-| <1$).

To sum up, Eqs.~(\ref{SM-eq-ev-7}, \ref{SM-eq-ev-99}, \ref{SM-eq-ev-9}) give the $N$ eigenvalues of the $\mathcal{S}_-$ operator. These results are summarized in Table~\ref{tab:Spm-eigs}, which also lists the eigenvalues of $\mathcal{S}_+$.

\begin{table}[t]
\centering
\renewcommand{\arraystretch}{1.25}
\setlength{\tabcolsep}{6pt}
\begin{tabular}{c c c c c}
\hline
Operator & Type & Eigenvalues & Count for even $N$ & Count for odd $N$ \\
\hline
$\mathcal{S}_{-}$ 
& $|\lambda_-|=1$ 
& $\{e^{-iE_k\tau}\}$ with $k$ odd 
& $N/2$ & $(N+1)/2$ \\[4pt]

$\mathcal{S}_{-}$ 
& $|\lambda_-|=0$ 
& $0$ 
& $1$ & $1$ \\[4pt]

$\mathcal{S}_{-}$ 
& $0<|\lambda_-|<1$ 
& \parbox{0.50\columnwidth}{\centering
$\sum_{k\in\mathrm{even}}
\frac{\sin^2(\omega_k)}{\lambda_{-}-e^{2i\tau\cos(\omega_k)}}=0$
}
& $N/2-1$ & $(N-3)/2$ \\
\hline

$\mathcal{S}_{+}$ 
& $|\lambda_+|=1$ 
& $\{e^{-iE_k\tau}\}$ with $k$ even 
& $N/2$ & $(N-1)/2$ \\[4pt]

$\mathcal{S}_{+}$ 
& $|\lambda_+|=0$ 
& $0$ 
& $1$ & $1$ \\[4pt]

$\mathcal{S}_{+}$ 
& $0<|\lambda_+|<1$ 
& \parbox{0.50\columnwidth}{\centering
$\sum_{k\in\mathrm{odd}}
\frac{\sin^2(\omega_k)}{\lambda_{+}-e^{2i\tau\cos(\omega_k)}}=0$
}
& $N/2-1$ & $(N-1)/2$ \\
\hline
\end{tabular}
\caption{Eigenvalues of the survival operators $\mathcal{S}_{\pm}=(\mathbbm{1}-|d_\pm\rangle\langle d_\pm|)U(\tau)$ for the nearest-neighbor tight binding Hamiltonian. The results hold provided that the quasienergies do not coincide, \textit{i.e.}, Eq.~\eqref{resonance-eqn} is not satisfied.}
\label{tab:Spm-eigs}
\end{table}

\section{Relation of the splitting probability with the eigenvalues of the survival operator}
\label{sec-SM-relation}
In the Letter, we pointed out that the interference term $\xi(x_0,N,\tau)$ admits two equivalent spectral expansions, one in terms of the eigenvalues $\{ \lambda_-\}$ of the survival operator $\mathcal{S}_-$ and the other in terms of $\{ \lambda_+\}$ of $\mathcal{S}_+$. Here, we derive both expansions. 
\subsection{$\xi(x_0,N,\tau)$ as an expansion in $\{ \lambda_-\}$}
Recall that the splitting probabilities are given by
\begin{align}
P_L(x _0) = \frac{1}{2}+ \xi (x _0, N, \tau),~~~ P_R(x _0) = \frac{1}{2}-\xi (x _0, N, \tau), \label{supp-letter-neq-2}
\end{align}
where the term $\xi(x_0, N, \tau)$ is given in terms of the generating functions $\widetilde{\chi}^{(\pm)}(z) $ in Eq.~\eqref{SP-SR-eq-6}. By contrast, using the other yet equivalent expression of $\xi(x_0, N, \tau)$ in Eq.~\eqref{SP-SR-eq-6-nn}, we yield a different expansion, as shown later.

The generating functions are given in Eq.~\eqref{SI-3j81}, and plugging them in Eq.~\eqref{SP-SR-eq-6} gives
\begin{align}
& \xi(x_0, N, \tau) = \text{Re} \left[ \oint_{|z|=1} dz~\mathcal{I}(z, x_0,N,\tau) \right], ~~\text{with} \label{supp-SI-4}\\
& \mathcal{I}(z,x_0,N, \tau) = \frac{1}{4 \pi  i z}  \left[ \frac{\sum _{k \in ~\rm{even}} \sin (\omega _k) \sin (\omega_k x_0 ) \left( \frac{e^{-2 i  \tau \cos (\omega _k)}}{z- ~e^{-2 i  \tau \cos (\omega _k)}}  \right)  }{   \sum _{k \in ~\rm{even}} \sin^2 (\omega _k)  \left( \frac{1}{z- ~e^{-2 i  \tau \cos (\omega _k)}}  \right)      } \right] \left[ \frac{\sum _{k \in ~\rm{odd}} \sin (\omega _k) \sin (\omega_k x_0 ) \left( \frac{e^{2 i  \tau \cos (\omega _k)}}{1-z ~e^{2 i \tau \cos (\omega _k)}}  \right)  }{   \sum _{k \in ~\rm{odd}} \sin ^2 (\omega _k)  \left( \frac{1}{1-z ~e^{2 i  \tau \cos (\omega _k)}}  \right)      } \right] , \label{supp-pole-eq-2}
\end{align}
where $\omega _k = \pi k /(N+1)$. Eq.~\eqref{supp-SI-4} indicates the contour integration of a complex function $\mathcal{I}(z, x_0, N, \tau)$ over a unit circle. By the residue theorem, this integral can be evaluated as the sum of the residues of $\mathcal{I}(z,x_0,N,\tau)$ at its singularities inside the contour. Let us analyze its singularities. 

A numerical inspection of Eq.~\eqref{supp-pole-eq-2} reveals that the only singularities present are simple poles. The first pole is $z_{\rm pole}=0$. Other poles are obtained by solving
\begin{align}
& \sum _{k \in ~\rm{even}} \sin^2 (\omega _k)  \left( \frac{1}{z_{\rm pole}- ~e^{-2 i  \tau \cos (\omega _k)}}  \right)  = 0, \label{suppSR-eq-pole-1} \\
&  \sum _{k \in ~\rm{odd}} \sin ^2 (\omega _k)  \left( \frac{1}{1-z_{\rm pole} ~e^{2 i  \tau \cos (\omega _k)}}  \right) =0. \label{suppSR-eq-pole-2}
\end{align}
Comparing Eq.~\eqref{suppSR-eq-pole-1} with the eigenvalue condition for $\mathcal{S}_-$ in Table~\ref{tab:Spm-eigs}, we identify $z_{\rm pole}=\lambda_-^{*}$ (where ${}^*$ denotes complex conjugation). In particular, Eq.~\eqref{suppSR-eq-pole-1} corresponds to the nontrivial eigenvalues of $\mathcal{S}_-$, which satisfy $0<|\lambda_-|<1$. Hence $0<|z_{\rm pole}|<1$, so these poles lie inside the unit disk and contribute to the contour integral in Eq.~\eqref{supp-SI-4}.

Similarly, comparing Eq.~\eqref{suppSR-eq-pole-2} with Table~\ref{tab:Spm-eigs}, we find that these poles are related to the eigenvalues of $\mathcal{S}_+$ via $z_{\rm pole}=1/\lambda_{+}$. Since Eq.~\eqref{suppSR-eq-pole-2} corresponds to the nontrivial eigenvalues of $\mathcal{S}_+$, which satisfy $0<|\lambda_+|<1$, it follows that $|z_{\rm pole}|=|1/\lambda_+|>1$. Hence these poles lie outside the unit disc and do not contribute to the contour integral in Eq.~\eqref{supp-SI-4}. 

Therefore, $\xi(x_0,N,\tau)$ in this particular representation is determined solely by the poles $z_{\rm pole}=\lambda_-^{*}$ corresponding to the  eigenvalues of $\mathcal{S}_-$ with $|\lambda_-|<1$ (including $\lambda_-=0$)
\begin{align}
\xi(x_0,N, \tau) = -2 \pi~   \sum_{ |\lambda_-|<1} \text{Im} \left[ \operatorname{Residue}\!\left[\mathcal{I};\, z=\lambda_-^*\right] \right], \label{pole-eq-4}
\end{align} 
where $\operatorname{Residue}\!\left[\mathcal{I};\, z=\lambda_-^*\right]$ denotes the residue of the function $\mathcal{I}( z, x_0,N,\tau)$ at the pole $z=\lambda_-^*$. This relation makes explicit that $\xi(x_0,N,\tau)$ can be written as a summed contribution of the eigenvalues $\{\lambda_-\}$. 

{\color{black}
\subsection{$\xi(x_0,N,\tau)$ as an expansion in $\{ \lambda_+\}$}
We now show that $\xi(x_0,N,\tau)$ can also be written as a summed contribution of the eigenvalues $\{\lambda_+\}$ of the survival operator $\mathcal{S}_+$. To show this, we begin with $\xi(x_0,N,\tau)$ in Eq.~\eqref{SP-SR-eq-6-nn} and rewrite it as
\begin{align}
& \xi(x_0, N, \tau) = \text{Re} \left[ \oint_{|z|=1} dz~\mathcal{J}(z, x_0,N,\tau) \right], ~~\text{with} \label{supp-SI-4-pssrk}\\
& \mathcal{J}(z,x_0,N, \tau) = \frac{1}{4 \pi  i z}  \left[ \frac{\sum _{k \in ~\rm{odd}} \sin (\omega _k) \sin (\omega_k x_0 ) \left( \frac{e^{-2 i  \tau \cos (\omega _k)}}{z- ~e^{-2 i  \tau \cos (\omega _k)}}  \right)  }{   \sum _{k \in ~\rm{odd}} \sin^2 (\omega _k)  \left( \frac{1}{z- ~e^{-2 i  \tau \cos (\omega _k)}}  \right)      } \right] \left[ \frac{\sum _{k \in ~\rm{even}} \sin (\omega _k) \sin (\omega_k x_0 ) \left( \frac{e^{2 i  \tau \cos (\omega _k)}}{1-z ~e^{2 i \tau \cos (\omega _k)}}  \right)  }{   \sum _{k \in ~\rm{even}} \sin ^2 (\omega _k)  \left( \frac{1}{1-z ~e^{2 i  \tau \cos (\omega _k)}}  \right)      } \right] , \label{supp-pole-eq-2-pssrk}
\end{align}
With this form, one can perform the same analysis as before to show that the only contributing poles to Eq.~\eqref{supp-SI-4-pssrk} are $z_{\rm pole} = \lambda_+^*$, while the other poles $z_{\rm pole} = 1/\lambda_-$ do not contribute. Following the same steps as before, we then obtain
\begin{align}
\xi(x_0,N, \tau) = -2 \pi~   \sum_{ |\lambda_+|<1} \text{Im} \left[ \operatorname{Residue}\!\left[\mathcal{J};\, z=\lambda_+^*\right] \right], \label{pole-eq-4-PSSRk}
\end{align}
Thus in this representation, only the $\{ \lambda_+ \}$ eigenvalues contribute to $\xi(x_0,N, \tau)$ and hence control the splitting probability.}

\section{Spectral origin of the transition in the Splitting probability}
\label{sec-SM-spectral}
\textcolor{black}{In the Letter, we demonstrated that for $N\gg 1$, the splitting probabilities exhibit a transition at a critical sampling time $\tau_c=\pi/2$. For $\tau\le \tau_c$, both $P_L(x_0)$ and $P_R(x_0)$ are equal to $1/2$, independent of $x_0$ and $\tau$. Above criticality, for $\tau>\tau_c$, they instead depend on $x_0$ and $\tau$. As was pointed out in the Letter, this behavior originates from a qualitative change in the arrangement of the eigenvalues $\{\lambda_{\pm}\}$ of the operators $\mathcal{S}_{\pm}$, which, according to Eqs.~(\ref{pole-eq-4}-\ref{pole-eq-4-PSSRk}), induces a change in the interference term $\xi(x_0,N,\tau)$ and hence the splitting probabilities via Eq.~\eqref{supp-letter-neq-2}. Let us first look at the arrangement of $\{\lambda_{-}\}$ in the complex plane.}


\begin{figure}[t]
	\centering
	\includegraphics[scale=0.2]{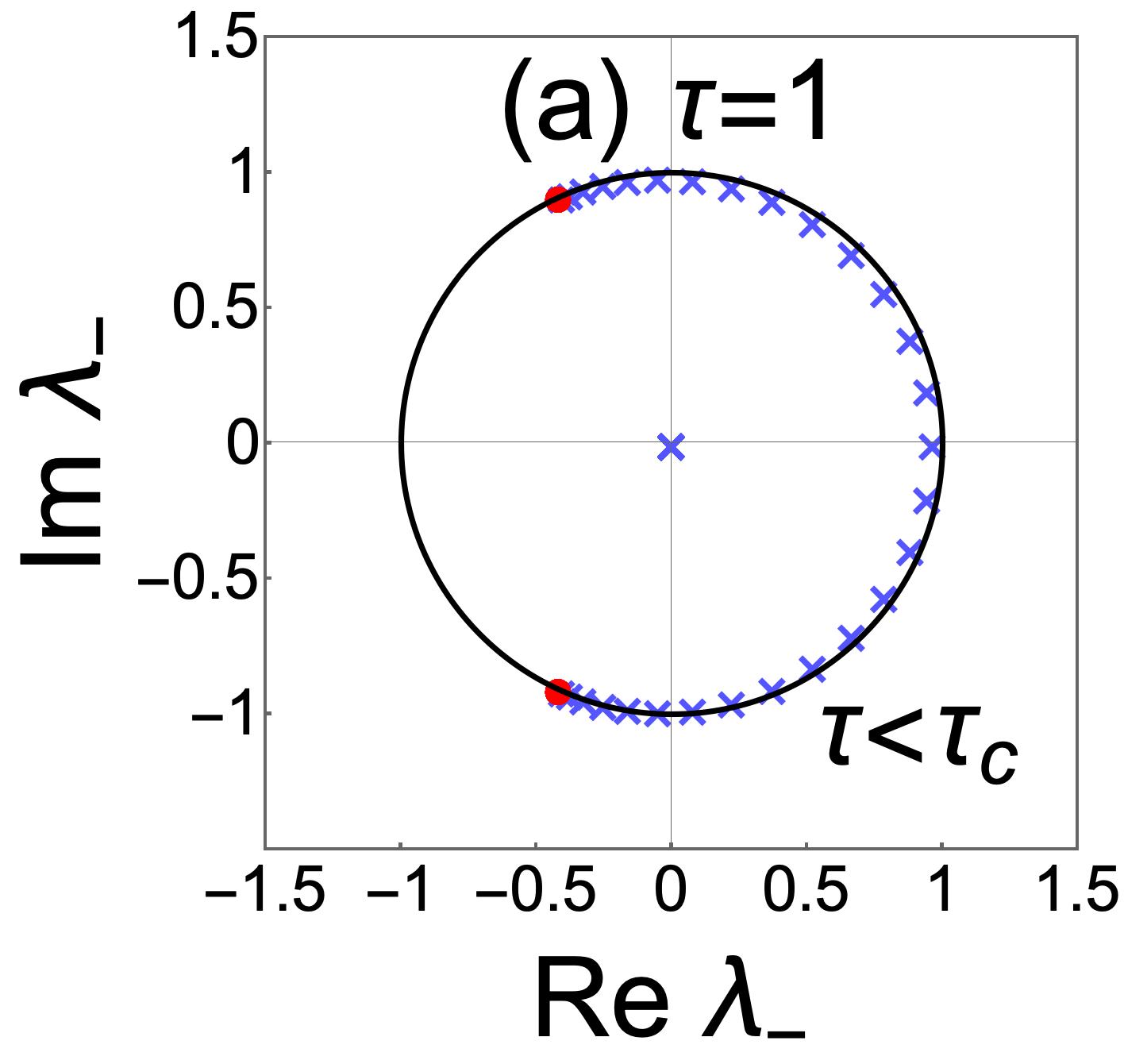}
	\includegraphics[scale=0.2]{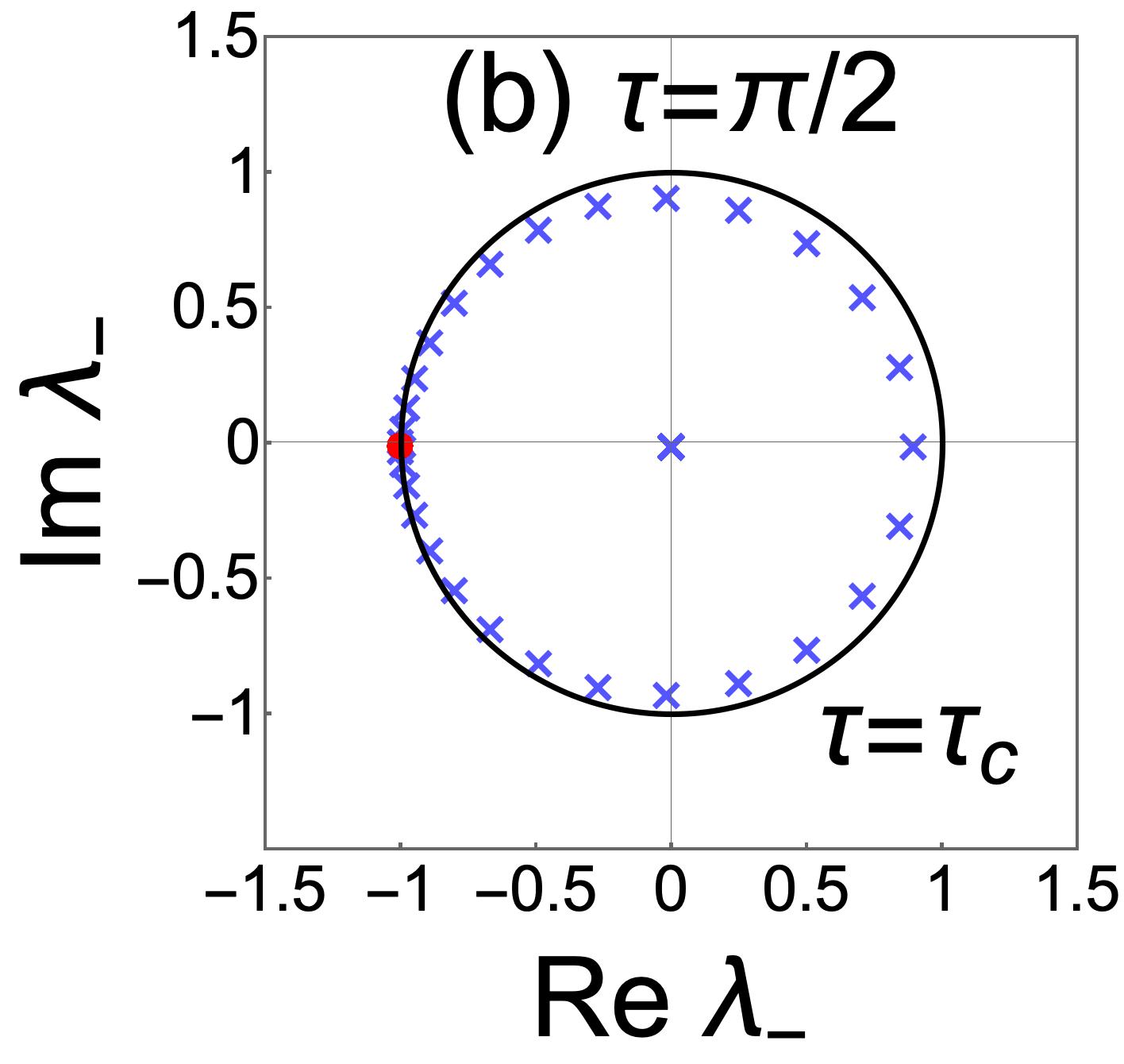}
	\includegraphics[scale=0.2]{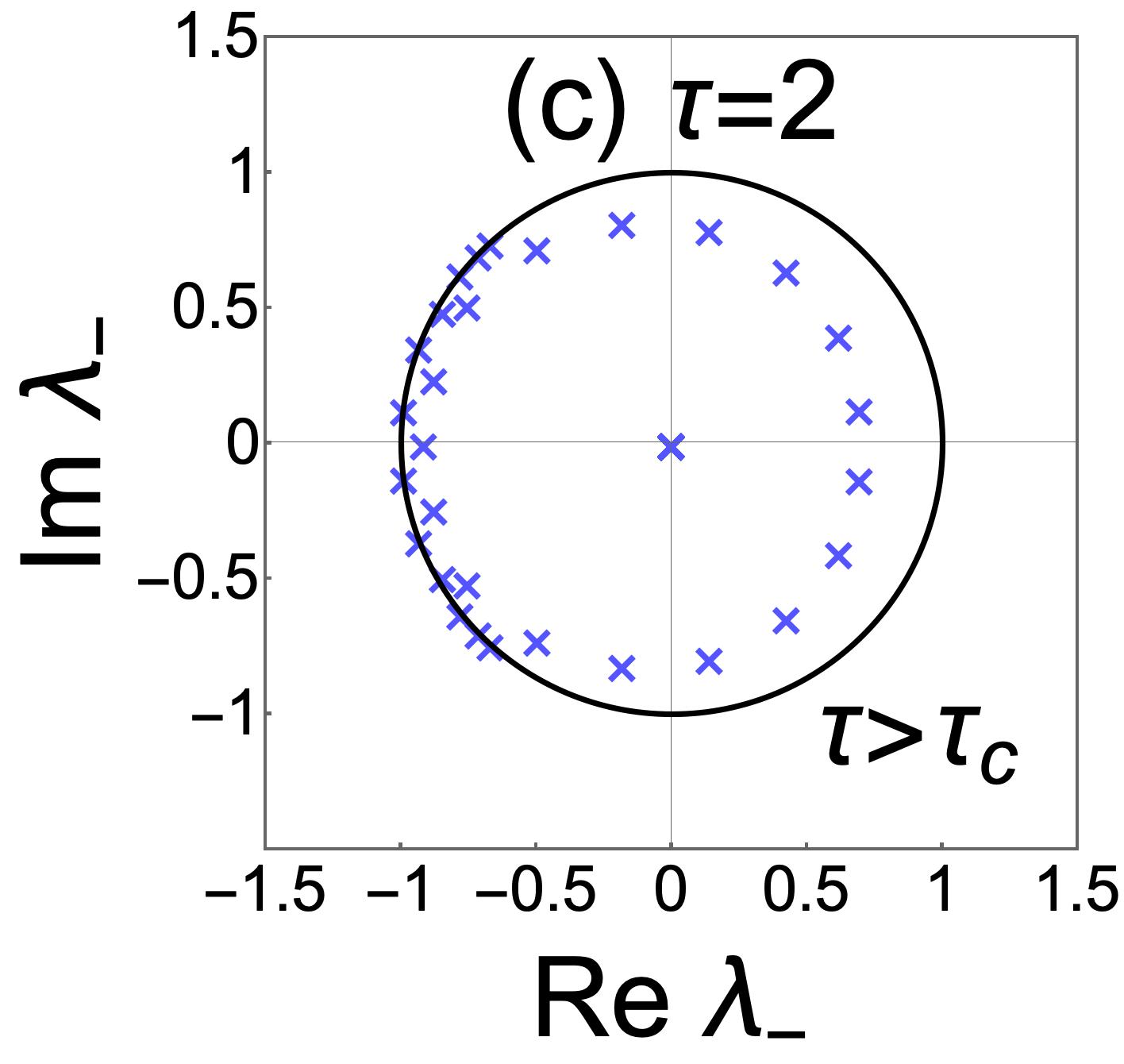}
	\includegraphics[scale=0.2]{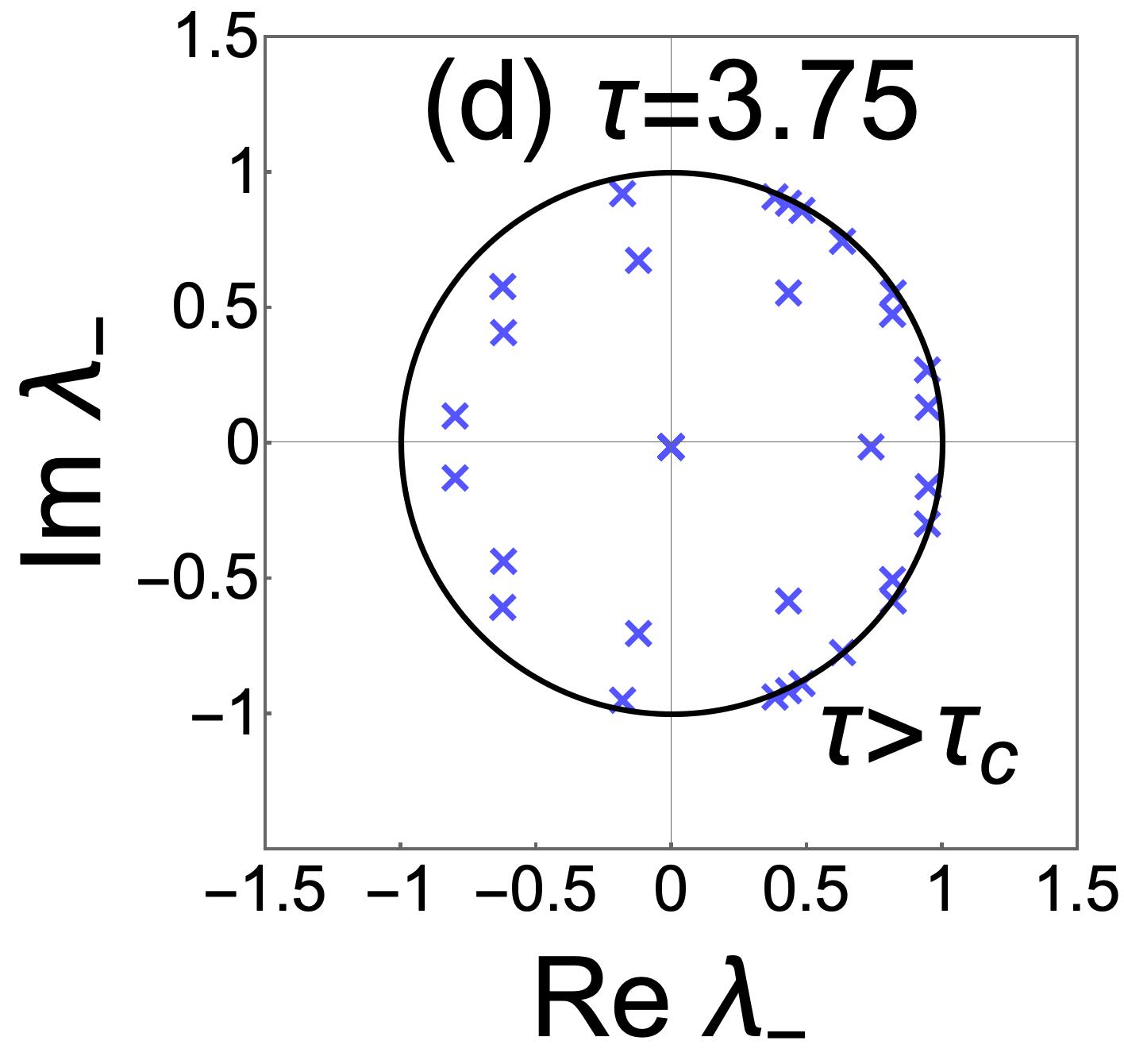}
	\includegraphics[scale=0.2]{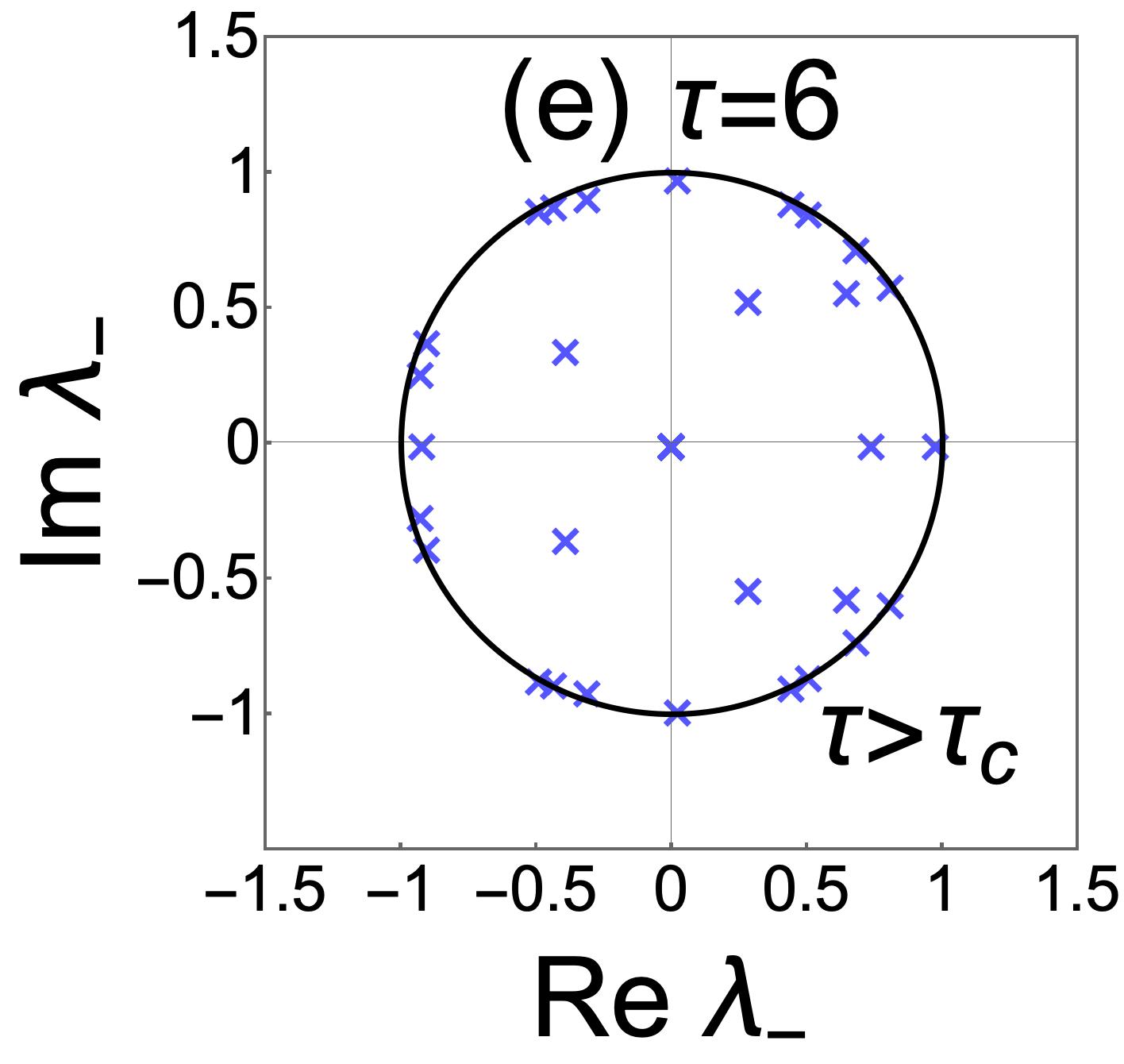}
	\caption{Eigenvalues $\{ \lambda_- \}$ (where $|\lambda_- <1|$) of the survival operator $\mathcal{S}_- = \left( 1-\ket{d_-} \bra{d_-}\right)U(\tau)$ for the nearest-neighbor tight-binding Hamiltonian. They are obtained by solving Eq.~\eqref{SM-eq-ev-9} numerically (also see Table~\ref{tab:Spm-eigs}). Panels~(a-e) show these eigenvalues for five different sampling times with $N=61$. For $\tau \leq \tau_c$, they are distributed along a single arc between the points $a_1(\tau) = e^{-iE_{\rm max} \tau}$ and $a_2(\tau) = e^{-iE_{\rm min} \tau}$ [shown in red in (a-b)] where $E_{\rm max} \approx 2$ and $E_{\rm min} \approx -2$ are the maximal and minimal energy eigenvalues for $N \gg 1$. For $\tau>\tau_c$, there is a qualitative change in the eigenvalue arrangement. The single arc breaks into distinct branches in the unit disk, and the arrangement becomes somewhat irregular. This, in turn, leads to qualitative change in $\xi(x_0,N, \tau)$ in Eq.~\eqref{pole-eq-4}. For $x_0=11,~N=61$, we find (a) $\xi(x_0,N,\tau) = 0.002$ for $\tau=1$, (b) $\xi(x_0,N,\tau) = -0.002$ for $\tau=\tau_c$, (c) $\xi(x_0,N,\tau) = -0.03$ for $\tau=2$, (d) $\xi(x_0,N,\tau) = -0.135$ for $\tau=3.75$ and (a) $\xi(x_0,N,\tau) = 0.132$ for $\tau=6$. Plugging these results in Eq.~\eqref{supp-letter-neq-2}, we obtain $P_L(x_0) \approx P_R(x_0) \approx 1/2$ for $\tau \leq \tau _c$ and $P_L(x_0) \neq P_R(x_0) \neq 1/2$ for $\tau > \tau _c$.}
	\label{fig-poles}
\end{figure}


In panels (a-e) of Fig.~\ref{fig-poles}, we employ Eq.~\eqref{SM-eq-ev-9} (also see Table~\ref{tab:Spm-eigs}) to numerically obtain the eigenvalues $\{ \lambda_- \}$ for five different sampling times, (a) $\tau = 1$, (b) $ \tau = \pi/2$, (c) $\tau = 2$, (d) $ \tau = 3.75$, and (e) $\tau = 6$. The system size is kept fixed to $N=61$. Panel (a) corresponds to $\tau < \tau_c$, panel (b) to $\tau = \tau_c$, and panels (c–e) to $\tau > \tau_c$. For $\tau = 1$, we see in (a) that all eigenvalues, except the one at the origin, lie close to the unit circle and are distributed regularly along a single arc between the points $a_1(\tau)$ and $a_2(\tau)$ (shown in red in (a)). By numerical inspection, we find $a_1(\tau)= \exp \left( -i E_{\rm max} \tau \right)$ and $a_2(\tau) = \exp \left(- i E_{\rm min} \tau \right)$, where $E_{\rm max} \approx 2$ and $E_{\rm min} \approx -2$ are the maximum and minimum energies for large system sizes. For such a regular arrangement of $\{\lambda_-\}$, the contributions associated with different eigenvalues in Eq.~\eqref{pole-eq-4} exhibit strong cancellations, resulting in a negligible $\xi(x_0,N,\tau)$. For example, with $x_0=11$ and $N=61$, we obtain $\xi(x_0,N, \tau=1) \approx 0.002$. Plugging this result in Eq.~\eqref{supp-letter-neq-2} then gives $P_{L}(x_0) \approx P_{R}(x_0) \approx 1/2$.


As we increase $\tau$, the eigenvalues remain still close to the unit circle, distributed regularly along a single arc between the points $a_1(\tau)$ and $a_2(\tau)$. However, as $\tau$ approaches the critical point
\begin{align}
\tau _c = 2 \pi / \Delta E 
\end{align}
where $\Delta E = (E_{\rm max}-E_{\rm min})$, the two points coalesce on the negative real axis, i.e., $a_1(\tau)=a_2(\tau)=-1$, and the arc closes; see the red points in (b). Here again, the term $\xi(x_0,N,\tau)$ in Eq.~\eqref{pole-eq-4} is small. For same values of $x_0$ and $N$, we find  $\xi(x_0,N, \tau=\pi/2) \approx -0.002$, leading to $P_{L}(x_0) \approx P_{R}(x_0) \approx 1/2$.

Beyond this critical point, the situation changes qualitatively. 
The points $a_1(\tau)$ and $a_2(\tau)$ move past each other on the unit circle. Here, we find that the arrangement of the $\{ \lambda_- \}$ changes and is no longer restricted to a single arc, see (c-e). Instead, the distribution becomes irregular,  with eigenvalues spreading nonuniformly inside the unit disk. Here, the contributions of different eigenvalues in Eq.~\eqref{pole-eq-4} no longer cancel as efficiently, and $\xi(x_0,N,\tau)$ becomes appreciable for $\tau>\tau_c$. For example, for $x_0=11$ and $N=61$, we obtain $\xi(x_0,N,\tau=2)\approx -0.03$, $\xi(x_0,N,\tau=3.75)\approx -0.135$, and $\xi(x_0,N,\tau=6)\approx 0.132$. Thus, the term $\xi(x_0, N, \tau)$ is not generally negligible for $\tau > \tau_c$. This leads to deviations from $1/2$ for $\tau > \tau _c$ for the splitting probabilities via Eq.~\eqref{supp-letter-neq-2}. 

To sum up, as $\tau$ is tuned across $\tau_c$, the eigenvalues $\{ \lambda_- \}$ arrangement undergoes a qualitative change, which directly alters the residue sum in Eq.~\eqref{pole-eq-4} and thereby leads to the transition in the splitting probabilities in Eq.~\eqref{supp-letter-neq-2}. \textcolor{black}{We demonstrate in Fig.~\ref{fig-poles-pp} that this transition can also be
understood from the equivalent spectral representation in Eq.~\eqref{pole-eq-4-PSSRk} in terms of the eigenvalues $\{ \lambda_+ \}$ which also exhibits a qualitative change in their arrangement in the complex plane.}

\begin{figure}[t]
	\centering
	\includegraphics[scale=0.2]{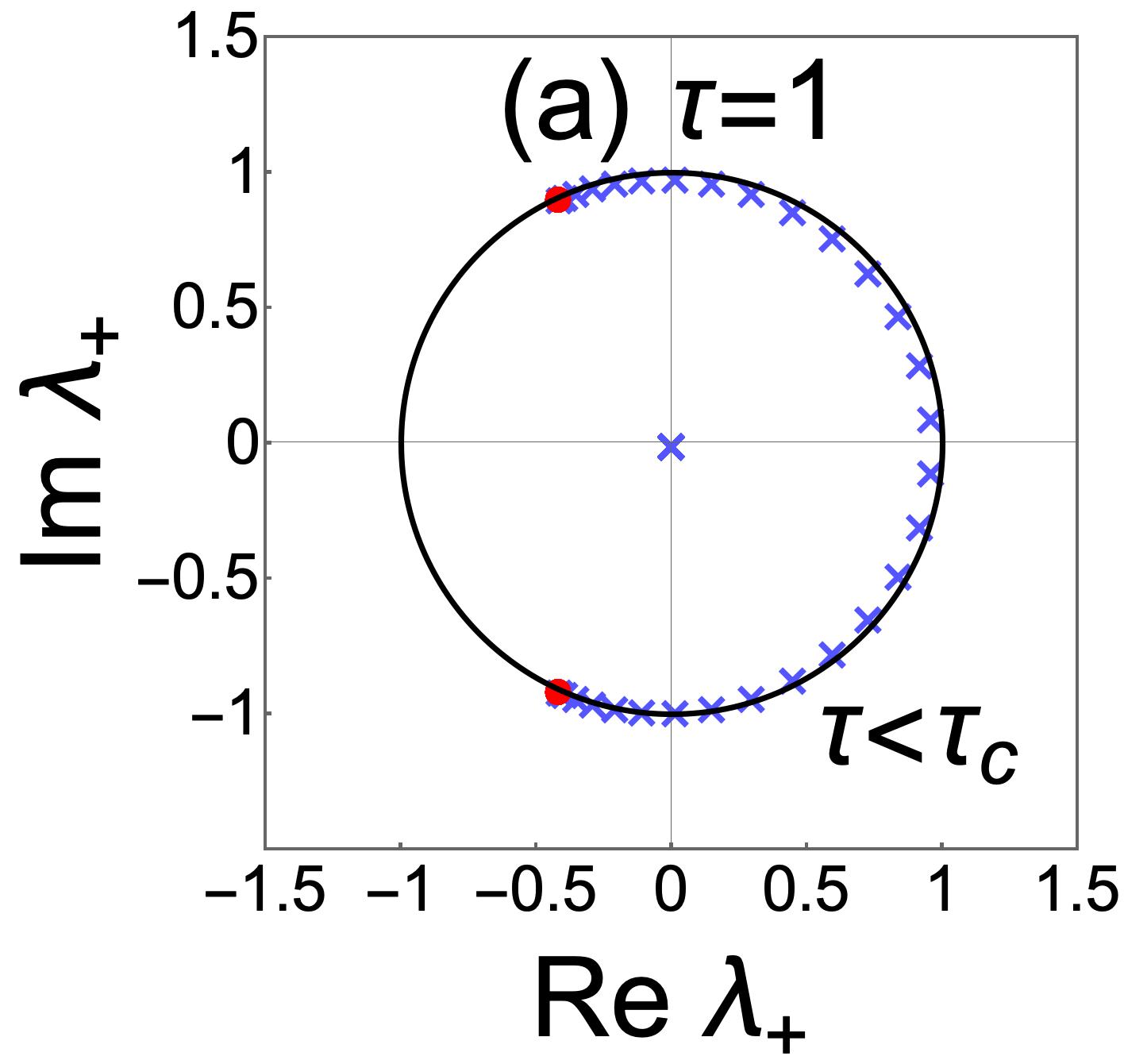}
	\includegraphics[scale=0.2]{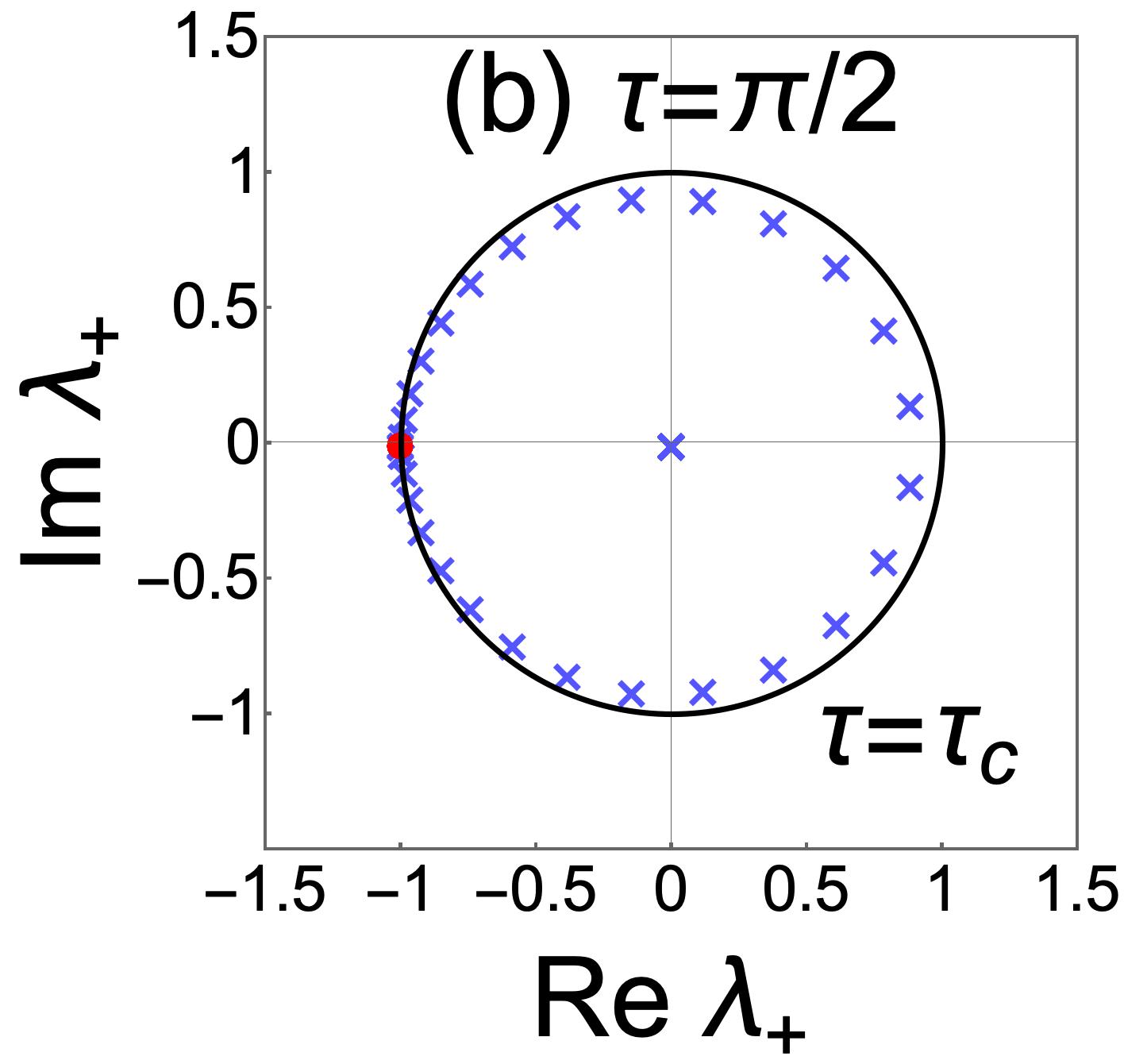}
	\includegraphics[scale=0.2]{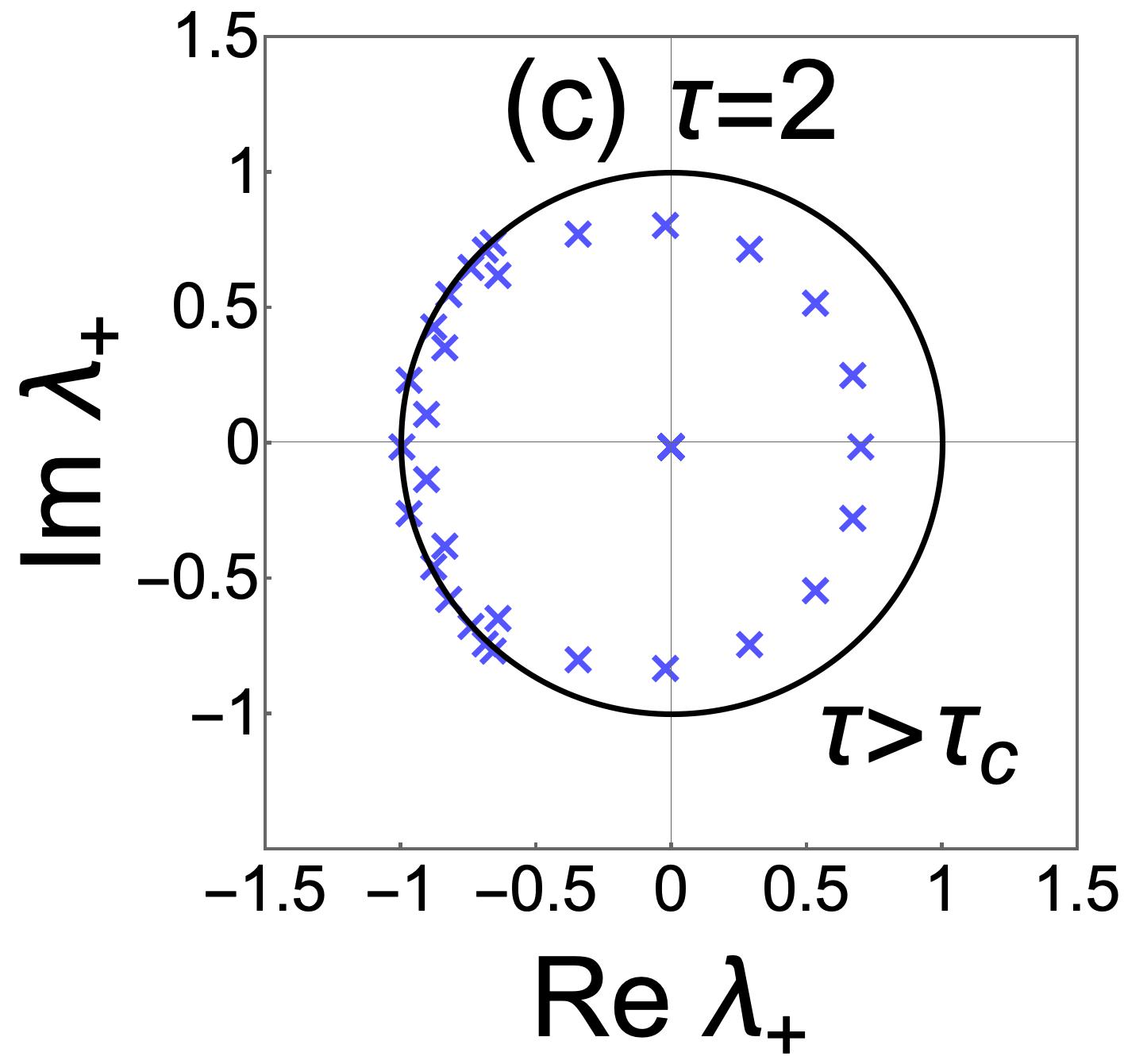}
	\includegraphics[scale=0.2]{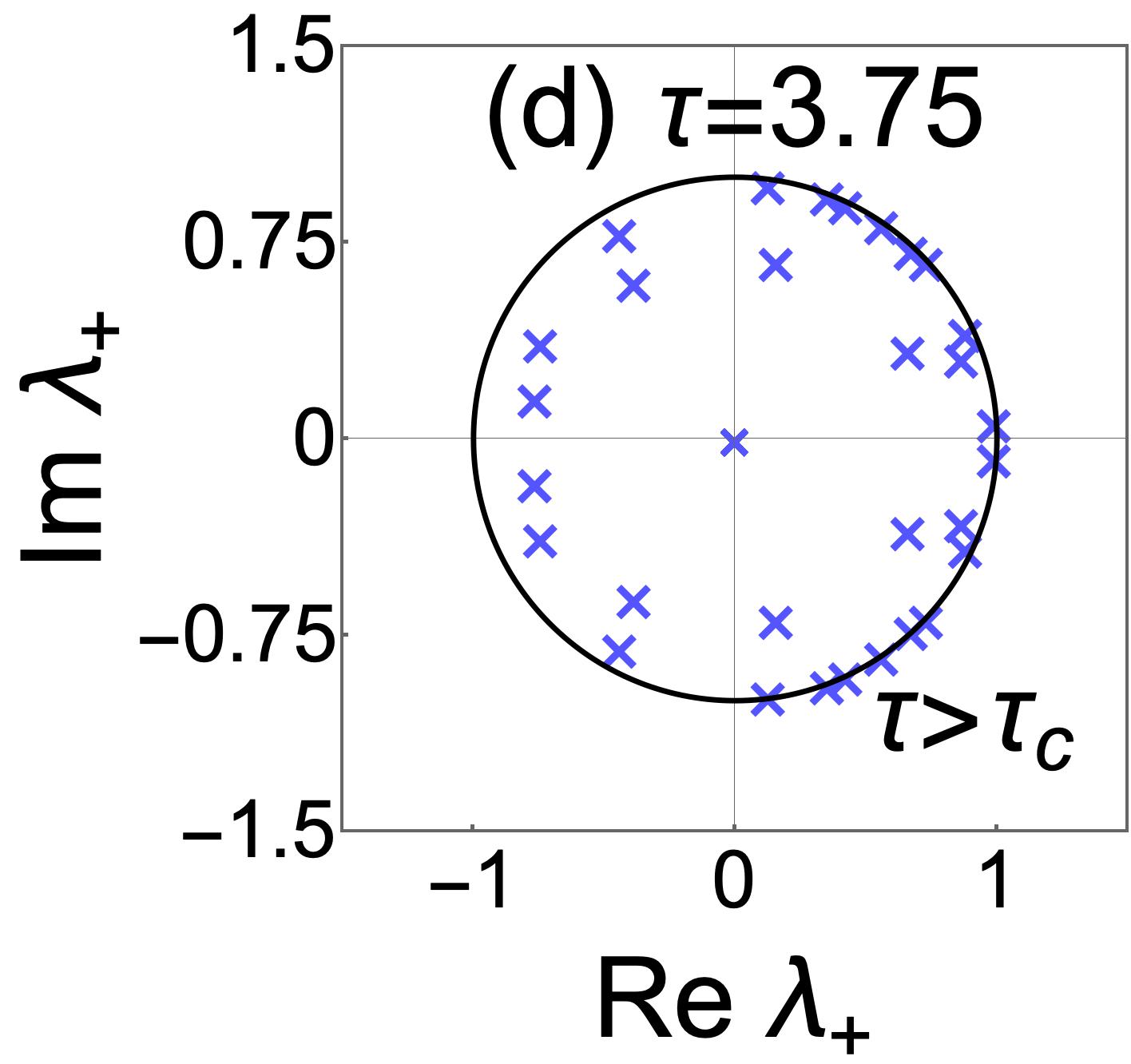}
	\includegraphics[scale=0.2]{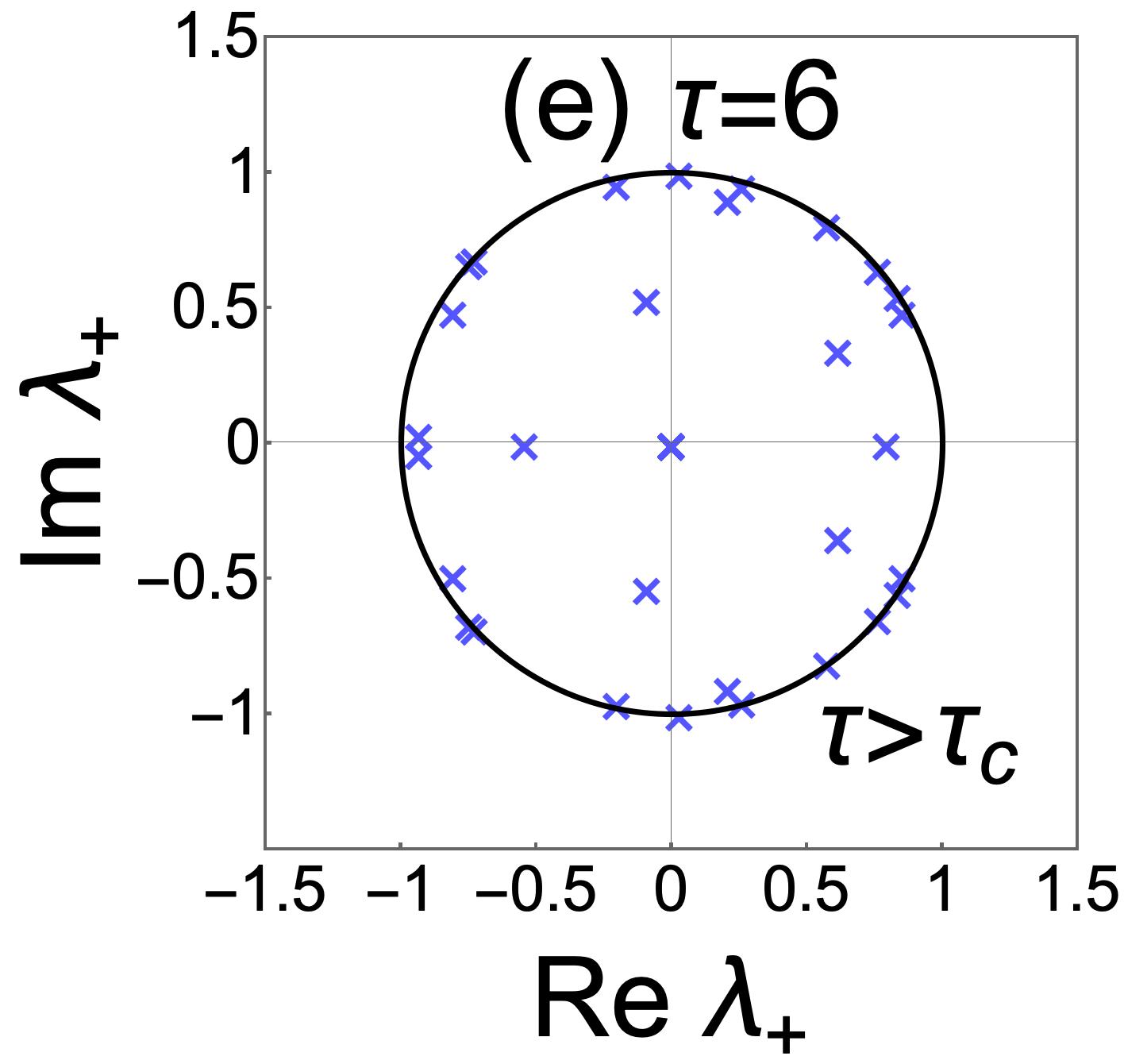}
	\caption{Demonstration of the transition in the arrangement of the eigenvalues $\{ \lambda_+ \}$  of the survival operator $\mathcal{S}_+ = \left( 1-\ket{d_+} \bra{d_+}\right)U(\tau)$ for the nearest-neighbor tight-binding Hamiltonian. They are obtained by solving the corresponding equation in Table~\ref{tab:Spm-eigs}. As done in Fig.~\ref{fig-poles}, we again choose five different sampling times with $N=61$ fixed for all cases. Once again, we see that for $\tau \leq \tau_c$, $\{ \lambda_+ \}$ are distributed along a single arc between the points $a_1(\tau) = e^{-iE_{\rm max} \tau}$ and $a_2(\tau) = e^{-iE_{\rm min} \tau}$ [shown in red in (a-b)] where $E_{\rm max} \approx 2$ and $E_{\rm min} \approx -2$ are the maximal and minimal energy eigenvalues for $N \gg 1$. For $\tau>\tau_c$, the arrangement becomes irregular with the single arc breaking into distinct branches in the unit disk. This reorganization produces the corresponding change in $\xi(x_0,N,\tau)$ through Eq.~\eqref{pole-eq-4-PSSRk}. For $x_0=11$ and $N=61$, we obtain $\xi(x_0,N,\tau)=0.002$ at $\tau=1$, $\xi(x_0,N,\tau)=-0.002$ at $\tau=\tau_c$, $\xi(x_0,N,\tau)=-0.03$ at $\tau=2$, $\xi(x_0,N,\tau)=-0.135$ at $\tau=3.75$, and $\xi(x_0,N,\tau)=0.132$ at $\tau=6$. Substitution into Eq.~\eqref{supp-letter-neq-2} then gives $P_L(x_0)\simeq P_R(x_0)\simeq 1/2$ for $\tau\leq\tau_c$, whereas for $\tau>\tau_c$ the splitting probabilities depart from the flat value and become unequal, $P_L(x_0)\neq P_R(x_0)$. Hence the transition in the splitting probability can also be understood in terms of the eigenvalues $\{ \lambda_+ \}$.}
	\label{fig-poles-pp}
\end{figure}

\begin{figure}[t]
	\centering
	\includegraphics[scale=0.3]{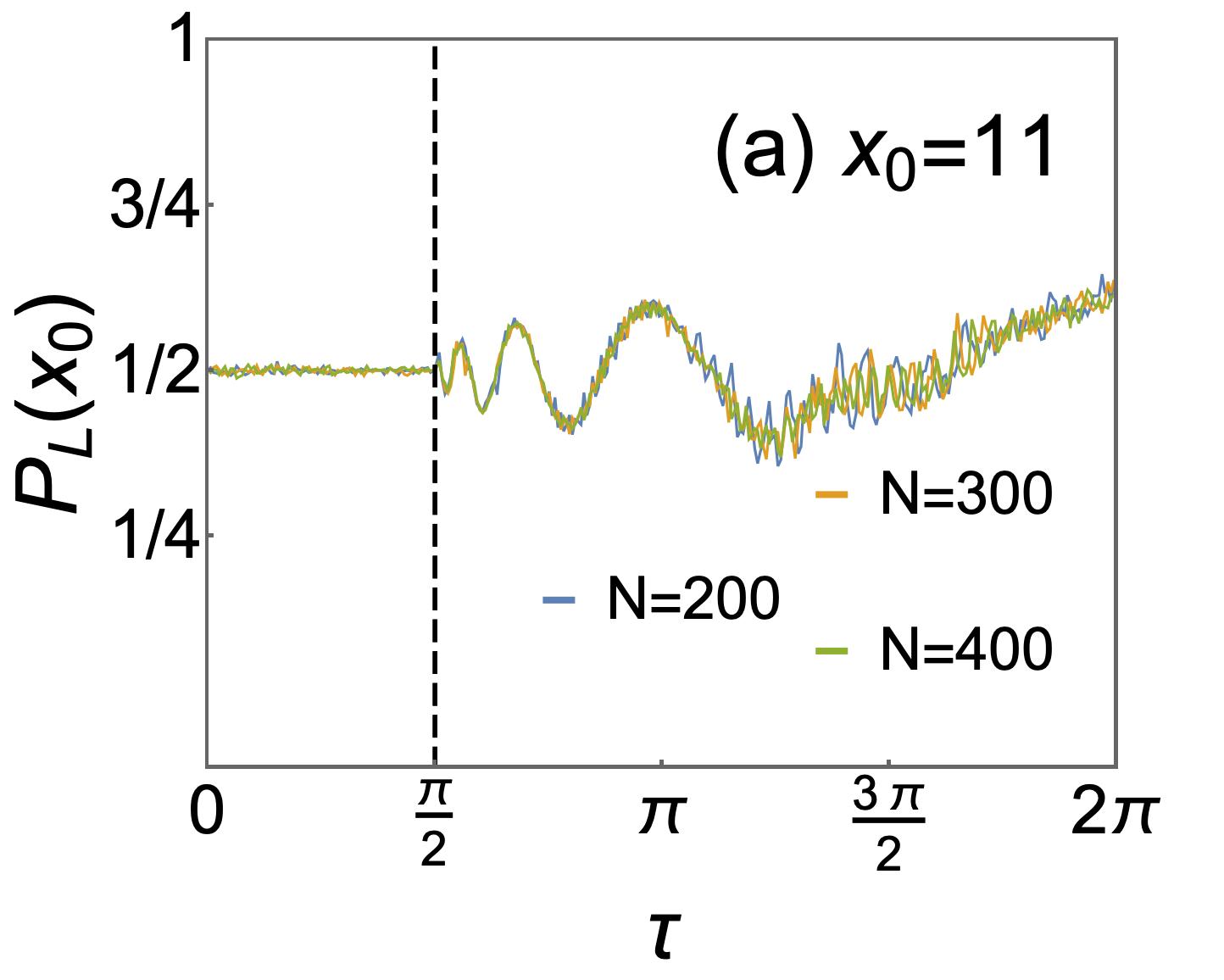}
	\includegraphics[scale=0.3]{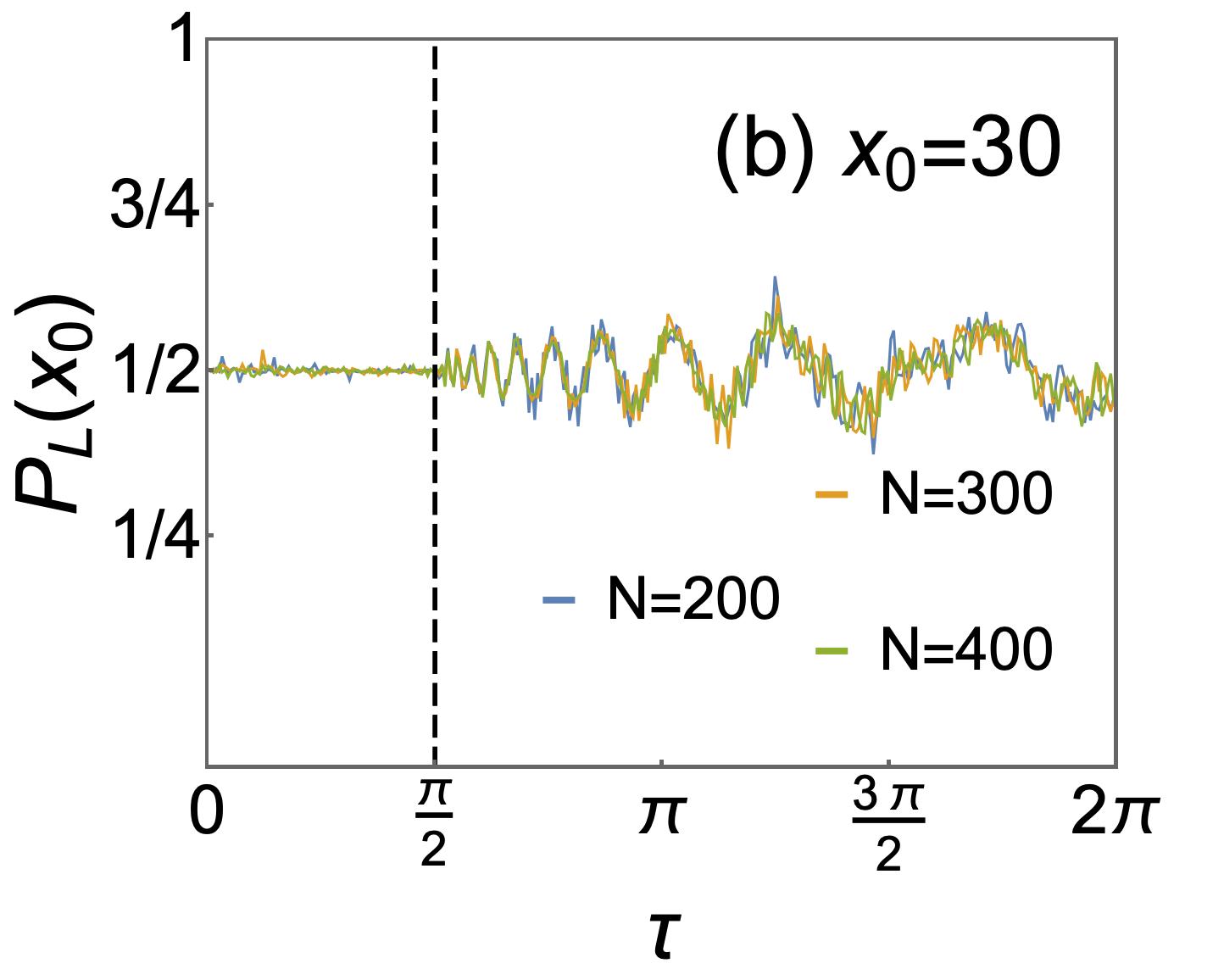} 
	\caption{Splitting probability is plotted as a function of the sampling time using Eq.~(6) of the Letter. Results are shown for two initial conditions: (a) $x_0=11$ and (b) $x_0=30$. For each case, three system sizes have been chosen, $N=200,~ 300, ~400$. In both panels, we observe data collapse for different values of $N$, indicating that the splitting probability is independent of the system size once it becomes very large ($N \gg 1$).}
	\label{fig-SM-pL-system-size}
\end{figure}

\section{Splitting probability is independent of $N$ when $N \gg 1$}
\label{appen-largeN}
{\color{black}
In the Letter, we plotted the splitting probability as a function of $\tau$ for one particular system size ($N=400$) and showed that it exhibits a transition at $\tau_c = 2\pi/\Delta E$. Here, we present the plot it for some other values of $N$. In Fig.~\ref{fig-SM-pL-system-size}, we show $P_L(x_0)$ as a function of $\tau$ for $x_0=11$ (left panel) and $x_0=30$ (right panel). In both panels, we have shown three system sizes, namely $N=200$ (blue), $N=300$ (yellow) and $N=400$ (green). We observe the data collapse across all panels, indicating that the probabilities are essentially independent of $N$ once the system size is sufficiently large.}

\section{Classical random walk under stroboscopic monitoring}
\label{appen-classical}
{\color{black}
To provide a meaningful comparison of our quantum results, we derive here the splitting probabilities of a classical random walk under stroboscopic monitoring. The continuous-monitoring $(\tau=0)$ case is a well-studied problem, and the splitting probabilities are given by the gambler's ruin formula 
\begin{align}
P_L^{\rm (cl)}(x_0) &= \frac{ \left( N - x_0 \right)}{\left( N - 1 \right)}, \qquad
P_R^{\rm (cl)}(x_0) = 1 - \left(\frac{N - x_0}{N - 1} \right).
\label{appen-unitary-eqn-classical}
\end{align}
As clearly seen, both probabilities are linear in $x_0$. However, to the best of our knowledge, the splitting probability for finite $\tau$ has not been investigated for the classical walks. 

Consider a continuous-time, discrete-space Markovian random walk consisting of a finite number of states $x \in \{ 1,2, \ldots, N \}$. We denote by $q^{(\mathrm{cl})}\!\left(x,t \mid x_0,0\right)$ the probability that the walker is at state $x$ at time $t$, given that it was initially at $x_0$. 
The superscript ``$\mathrm{cl}$", as indicated in the Letter, is used to emphasize the classical nature of the process. In the absence of any monitoring, the dynamics of the walker is governed by the master equation with transition matrix $\mathcal{W}$
\begin{align}
\frac{d}{dt} \ket{\mathcal{P}^{(\rm cl)} (t)  }= \mathcal{W}\ket{\mathcal{P}^{(\rm cl)} (0)}. \label{new-classical-pss-eq-1}
\end{align}
Here, we have defined the probability vector $\ket{\mathcal{P}^{(\rm cl)} (t)  }$ as
\begin{align}
 \ket{\mathcal{P}^{(\rm cl)} (t)  }  = \Big( q^{(\mathrm{cl})}\!\left(1,t \mid x_0,0\right),~~q^{(\mathrm{cl})}\!\left(2,t \mid x_0,0\right) , \ldots \ldots, q^{(\mathrm{cl})}\!\left(N,t \mid x_0,0\right)   \Big)^{T}, \label{new-classical-pss-eq-2}
\end{align}
($T$ for transpose). We designate two distinct target sites, $x = x_L$ and $x = x_R$, and monitor whether the walker is at either of these sites at the stroboscopic times $\tau,~ 2\tau, ~3\tau, \ldots$. This means that if the walker is located at these two sites at the measurement times, it is absorbed and the process is stopped. Otherwise, it continues to evolve until a successful detection is made. Therefore, in a given realization of the motion, the walker, starting from $x_0$, is evolved until it is detected for the first time at one of the designated target sites during a stroboscopic monitoring time (say at time $n_F \tau$). We then define $F_{\alpha}^{(\rm cl)} \left( n_F|x_0 \right)$ as the probability that the walker is first detected at $x_{\alpha}$ (with $\alpha \in \{L,R \}$ ) at time $n_F \tau$.

The splitting probabilties are given by
\begin{align}
P_L^{(\rm cl)} (x_0)= \sum _{n_F=1}^{\infty}  F_{L}^{(\rm cl)} \left( n_F|x_0 \right),~~P_R^{(\rm cl)}(x_0) = \sum _{n_F=1}^{\infty}  F_{R}^{(\rm cl)} \left( n_F|x_0 \right). \label{new-classical-pss-eq-3}
\end{align}
For a generic $\tau$, a renewal framework can be formulated which relates $F^{(\mathrm{cl})}_{\alpha}(n_F|x_0)$ to the detection-free propagator $q^{(\mathrm{cl})}(x,t|x_0,0)$ as follows
\begin{align}
F^{(\mathrm{cl})}_{\alpha}(n|x_0) = q^{(\mathrm{cl})}(x_{\alpha},n \tau|x_0,0) -  \sum _{\beta = L,R}~ \sum _{l=1}^{n-1} F^{(\mathrm{cl})}_{\beta}(l|x_0) ~q^{(\mathrm{cl})}(x_{\alpha},(n-l) \tau| x_{\beta},0) . \label{new-classical-pss-eq-4}  
\end{align}
The physical rationale behind this relation is as follows. The probability of a first detection at $x_\alpha$ after $n$ monitoring steps, without being detected at $x_{\bar{\alpha}}$ (where $\bar{\alpha}=R$ if $\alpha=L$ and vice versa), is obtained by taking the total probability of being at $x_\alpha$ at time $n\tau$ in the absence of monitoring, and subtracting all events in which the walker is detected by either detector, say $x_{\beta}$ with $\beta \in \{L, R\}$, at an earlier monitoring time $l\tau$, and subsequently propagates from $x_{\beta}$ to $x_\alpha$ in the remaining monitoring steps $(n-l)\tau$.

 Summing Eq.~\eqref{new-classical-pss-eq-4} over all $n$ yields splitting probabilities as
\renewcommand{\arraystretch}{1.5} 
\begin{align}
\begin{pmatrix}
1+\tilde{q}^{(\rm cl)}_{x_L, x_L} & \quad \tilde{q}^{(\rm cl)}_{x_L, x_R} \\
\tilde{q}^{(\rm cl)}_{x_R, x_L}   & \quad 1+\tilde{q}^{(\rm cl)}_{x_R, x_R}
\end{pmatrix}
\begin{pmatrix}
P_L^{(\rm cl)}(x_0) \\
P_R^{(\rm cl)}(x_0) 
\end{pmatrix}
=
\begin{pmatrix}
\tilde{q}^{(\rm cl)}_{x_L, x_0} \\
\tilde{q}^{(\rm cl)}_{x_L, x_0}
\end{pmatrix},
\label{new-classical-pss-eq-6}
\end{align}
\renewcommand{\arraystretch}{1} 
where
\begin{align}
\tilde{q}^{(\rm cl)}_{x, x_0} = \sum _{n=1}^{\infty}  q^{(\mathrm{cl})}(x,n\tau|x_0,0) . \label{new-classical-pss-eq-7}  
\end{align}

\begin{figure}[t]
	\centering
	\includegraphics[scale=0.3]{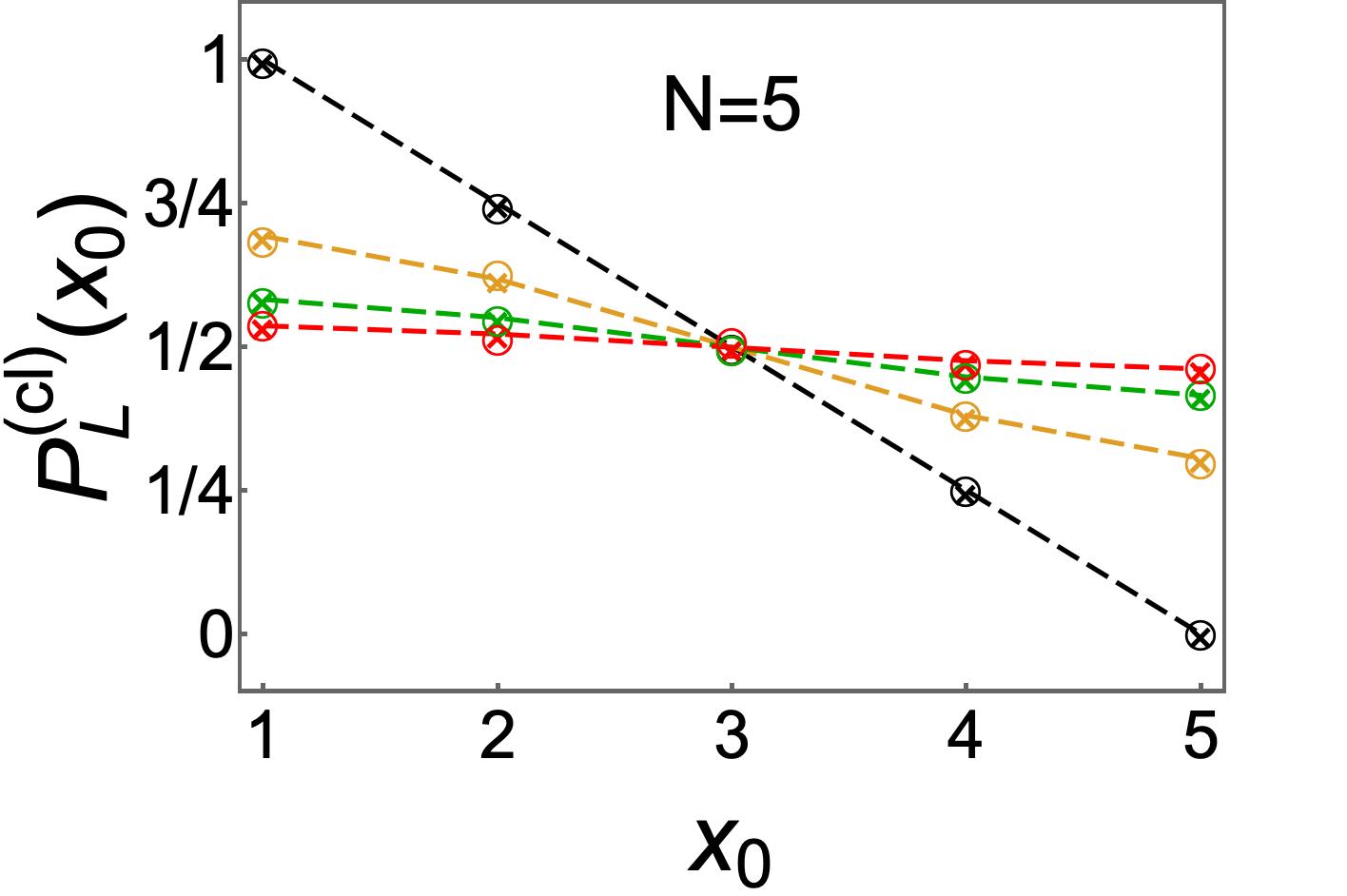} 
	\includegraphics[scale=0.3]{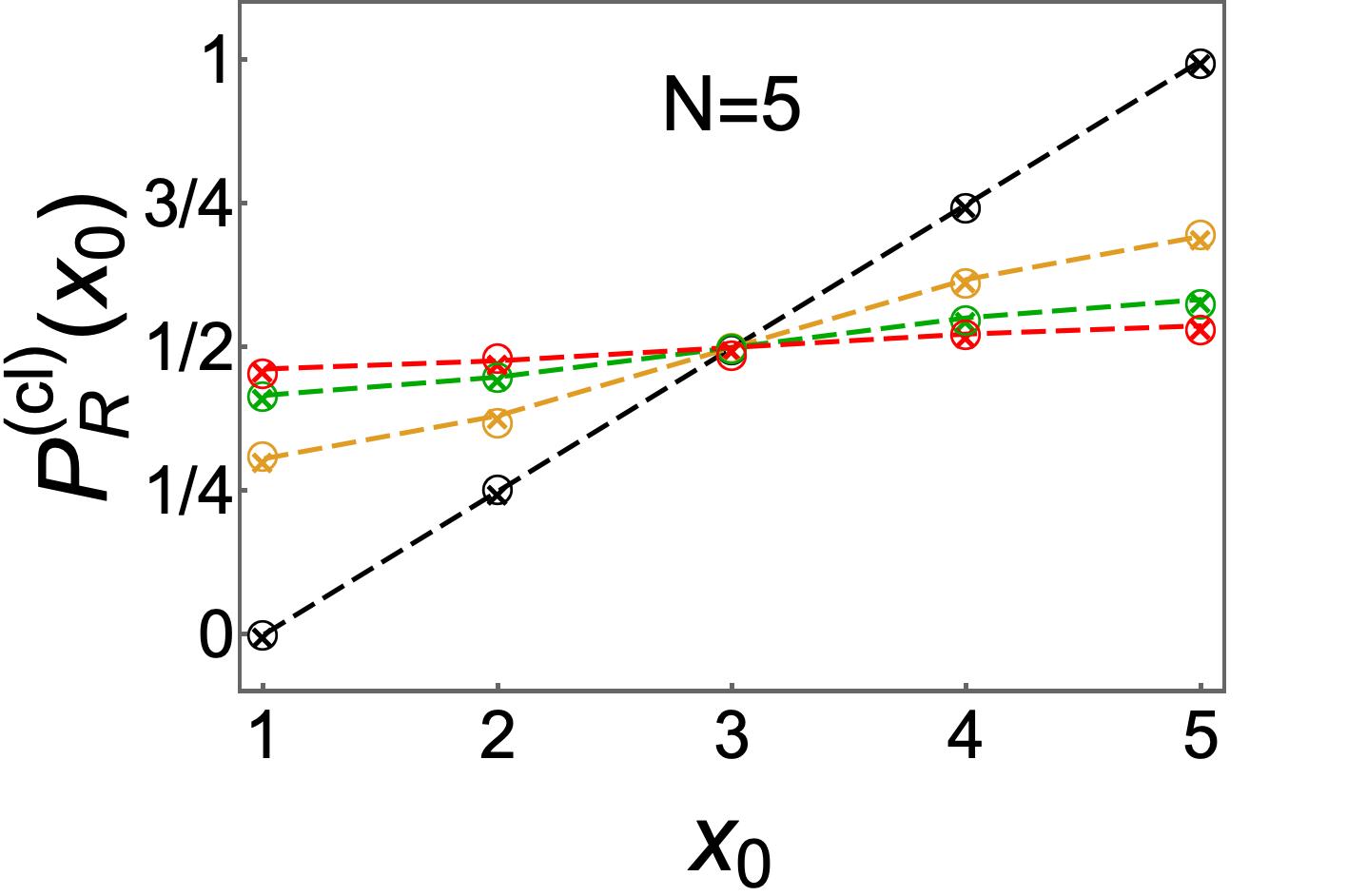}
	\caption{Splitting probabilities $P_L^{(\rm cl)}(x_0)$ and $P_R^{(\rm cl)}(x_0)$ for a classical random walk on a finite chain of size $N=5$ and monitoring at the endpoints, $x_L=1$ and $x_R=5$. In both panels, different colors represent different monitoring intervals: $\gamma \tau = 0$ (black), $\gamma \tau =2$ (yellow), $\gamma \tau = 4$ (green) and $\gamma \tau = 6$ (red). Furthermore, crosses $(\times)$ denote the numerical simulations, while circles $(\circ)$ indicate the theoretical predictions from Eqs.~\eqref{new-classical-pss-eq-13} and \eqref{new-classical-pss-eq-14} for $P_L^{(\rm cl)}(x_0)$ and $P_R^{(\rm cl)}(x_0)$, respectively. The splitting probabilities are defined only for integer values of $x_0$; dashed lines between points are included merely as a visual aid.
	}
	\label{fig-simulation-classical}
\end{figure}

\begin{figure}[t]
	\centering
	\includegraphics[scale=0.4]{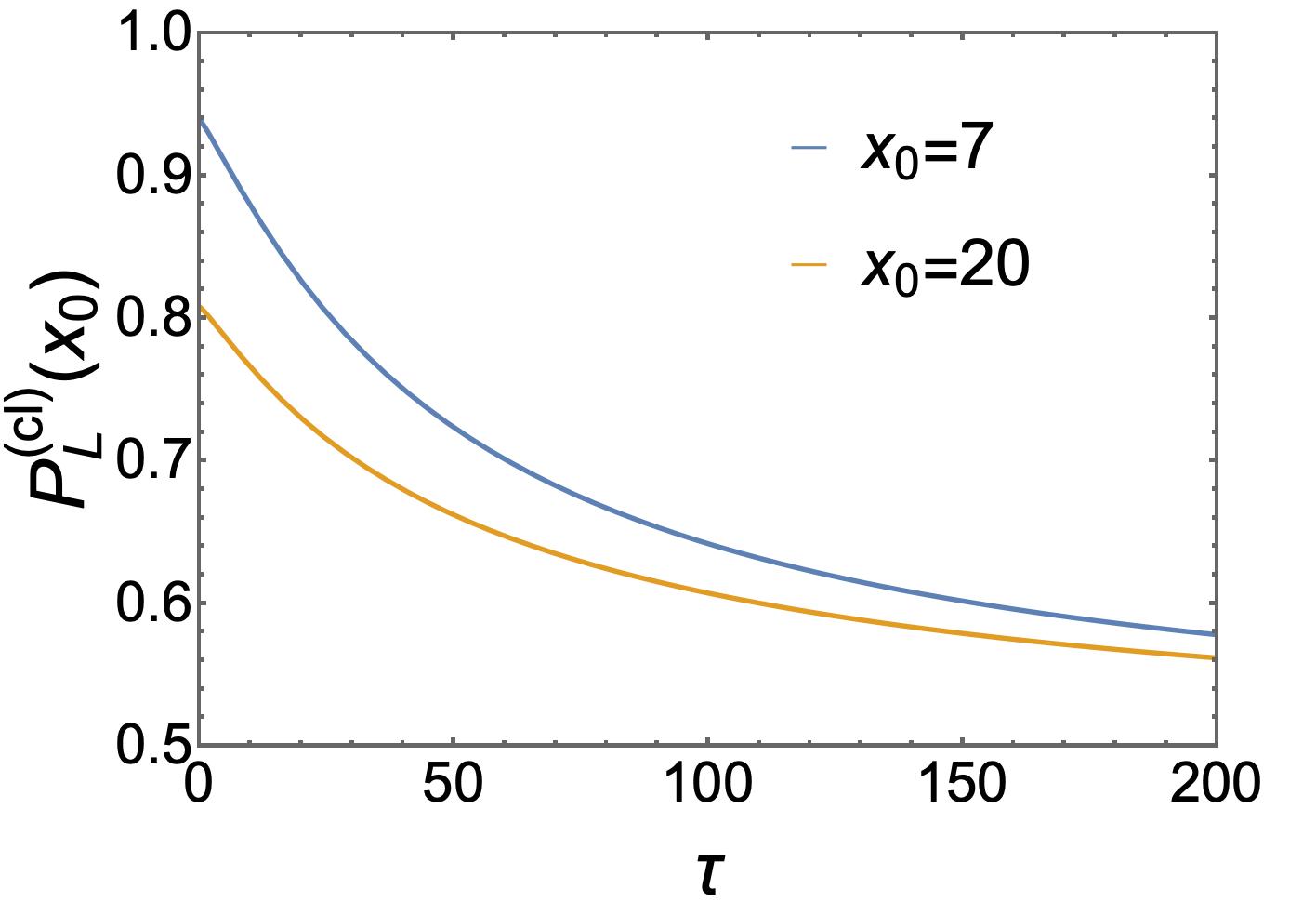} 
	\caption{Splitting probability of a classical random walk under stroboscopic monitoring is plotted as a function of the sampling time with $\gamma=1$ using Eq.~\eqref{new-classical-pss-eq-13}. Results are shown for $x_0=7$ and $x_0=20$, with the system size fixed at $N=50$ in both cases. In contrast to the quantum case, no phase-transition-like behavior is observed for the classical case.}
	\label{fig-classical}
\end{figure}

\noindent
Let us now consider the random walk in a one-dimensional segment of size $N$ with nearest neighbor jumps, in analogy with the nearest-neighbor quantum tight-binding Hamiltonian studied in the Letter. We label the lattice sites by $x=1,2, \ldots,N$. In the bulk, the walker can jump to one of its neighboring site with a rate $\gamma$. At the boundaries, these rules are modified: from site $x=1$, the walker can only jump to site $x=2$ with rate $\gamma$, while from site $x=N$, it can only jump to site $x=(N-1)$ with the same rate (similar to the quantum case). The transition matrix reads as
\begin{align}
\mathcal{W} = 
\begin{pmatrix}
-\gamma & \gamma   & 0        & 0        & \cdots & 0 \\
\gamma  & -2\gamma & \gamma   & 0        & \cdots & 0 \\
0       & \gamma   & -2\gamma & \gamma   & \cdots & 0 \\
0       & 0        & \gamma   & -2\gamma & \ddots & \vdots \\
\vdots  & \ddots   & \ddots   & \ddots   & \ddots & \gamma \\
0       & \cdots   & 0        & 0        & \gamma & -\gamma
\end{pmatrix}.
\label{new-classical-pss-eq-8}  
\end{align}
The eigenvalues are
\begin{align}
\mathcal{E} _  k = - 4 \gamma \sin ^2 \left( \frac{\pi k}{2 N} \right), \label{new-classical-pss-eq-9}  
\end{align}
$\text{where } k = 0,1,2, \ldots, (N-1)$. The $x$-component of the eigenvector $\ket{\mathcal{E}  _k}$ is
\begin{align}
\braket{x \big | \mathcal{E}  _k} = \sqrt{\frac{2 - \delta _{k,0}}{N}}~\cos \left[ \frac{k \pi}{N} \left( x - \frac{1}{2} \right) \right]. \label{new-classical-pss-eq-10}  
\end{align}
With this spectral-expansion, we can now write the free propagator $q^{(\mathrm{cl})}(x,n\tau|x_0,0) = \braket{x|e^{\mathcal{W}n \tau}|x_0}$ as
\begin{align}
q^{(\mathrm{cl})}(x,n\tau|x_0,0) = \sum _{k=0}^{N-1}   \left( \frac{2 - \delta _{k,0}}{N} \right) ~\cos \left[ \frac{k \pi}{N} \left( x - \frac{1}{2} \right) \right]~\cos \left[ \frac{k \pi}{N} \left( x_0 - \frac{1}{2} \right) \right]  e^{\mathcal{E} _k n \tau } , \label{new-classical-pss-eq-11} 
\end{align}
and inserting it in Eq.~\eqref{new-classical-pss-eq-7}, we obtain
\begin{align}
\tilde{q}^{(\rm cl)}_{x, x_0} = \sum _{k=0}^{N-1}   \left( \frac{2 - \delta _{k,0}}{N} \right) ~\cos \left[ \frac{k \pi}{N} \left( x - \frac{1}{2} \right) \right]~\cos \left[ \frac{k \pi}{N} \left( x_0 - \frac{1}{2} \right) \right] \frac{e^{\mathcal{E} _k \tau }}{1-e^{\mathcal{E} _k \tau }}. \label{new-classical-pss-eq-12} 
\end{align}
Eq.~\eqref{new-classical-pss-eq-6} then gives the splitting probabilities
\begin{align}
& P_L^{(\rm cl)}(x_0) = \frac{1}{2}+ \frac{2 \sum _{k=1, \rm odd}^{N-1} ~ \cos \left( \frac{\pi k}{2N} \right)~\cos \left( \frac{\pi k}{N} \left( x_0-\frac{1}{2} \right) \right) \left( \frac{e^{\mathcal{E} _k \tau }}{1-e^{\mathcal{E} _k \tau }} \right) }{N + 4 \sum _{k=1, \rm odd}^{N-1} ~ \cos \left( \frac{\pi k}{2N} \right)^2 \left( \frac{e^{\mathcal{E} _k \tau }}{1-e^{\mathcal{E} _k \tau }} \right)},  \label{new-classical-pss-eq-13}   \\
& P_R^{(\rm cl)}(x_0) = \frac{1}{2}- \frac{2 \sum _{k=1, \rm odd}^{N-1} ~ \cos \left( \frac{\pi k}{2N} \right)~\cos \left( \frac{\pi k}{N} \left( x_0-\frac{1}{2} \right) \right) \left( \frac{e^{\mathcal{E} _k \tau }}{1-e^{\mathcal{E} _k \tau }} \right) }{N + 4 \sum _{k=1, \rm odd}^{N-1} ~ \cos \left( \frac{\pi k}{2N} \right)^2 \left( \frac{e^{\mathcal{E} _k \tau }}{1-e^{\mathcal{E} _k \tau }} \right)}. \label{new-classical-pss-eq-14} 
\end{align}
These expressions provide the exact splitting probabilities of a classical random walker for arbitrary monitoring interval $\tau$ at $x_L=1$ and $x_R=N$, and for any finite system size $N$. For all values of $\tau$, we see that $P_L^{(\rm cl)}(x_0) + P_R^{(\rm cl)}(x_0)=1$. This is in sharp contrast to the quantum case, where at special sampling times we find $P_L(x_0)+P_R(x_0)<1$.

Fig.~\ref{fig-simulation-classical} demonstrates a comparison of our theoretical expressions with the numerical simulations for $N=5$, and four different choices of the monitoring interval: $\gamma \tau = 0$ (black), $\gamma \tau =2$ (yellow), $\gamma \tau = 4$ (green) and $\gamma \tau = 6$ (red). In both panels, the simulation data are shown by crosses $(\times)$, while the theoretical predictions are indicated by circles $(\circ)$. Our predictions in Eqs.~\eqref{new-classical-pss-eq-13} and \eqref{new-classical-pss-eq-13}, as expected, show excellent agreement with the numerical simulations for all parameter sets.

Let us examine the $P_{L}^{(\rm cl)}(x_0)$ plot in the left panel. For $\gamma \tau = 0$ (black), the plot is linear in $x_0$, consistent with the gambler's ruin formula in Eq.~\eqref{appen-unitary-eqn-classical}. In this limit, $P_L^{(\rm cl)}(x_0=1)=1$ and $P_L^{(\rm cl)}(x_0=N)=0$, reflecting the fact that with continuous monitoring, the walker is certainly absorbed at the left boundary if it starts there, and never reaches the right boundary. For intermediate values of $x_0$, the splitting probability satisfies the inequality relation
\begin{align}
\begin{cases}
P_L^{\rm (cl)}(x_0) \geq P_R^{\rm (cl)}(x_0)  , & ~~~\text{for }  x_0 \leq 3, \\[4pt]
P_L^{\rm (cl)}(x_0) \leq P_R^{\rm (cl)}(x_0)  , & ~~~\text{for }  x_0 > 3.
\end{cases}
\label{unitary-eqn-46}
\end{align}
This behavior follows because proximity to a detector enhances its splitting probability in that direction for the classical walker (similarly for $P_R^{\rm (cl)}(x_0)$). On increasing the stroboscopic interval $\tau$, $P_L^{\rm (cl)}(x_0)$ and $P_R^{\rm (cl)}(x_0)$ are no longer linear in $x_0$, see $\gamma \tau = 2,~4,$ in Fig.~\ref{fig-simulation-classical}. However, the inequality relation \eqref{unitary-eqn-46} continues to hold. Finally, for sufficiently large $\tau$, the two probabilities take a symmetric form, $P_L^{\rm (cl)}(x_0) \simeq P_R^{\rm (cl)}(x_0) \simeq \tfrac{1}{2}$, independent of $x_0$, see $\gamma \tau = 6$ (red) plot. This saturation at $1/2$ arises because, for $\gamma \tau \gg 1$, the walker equilibrates, and effectively loses memory of its initial position.

The classical splitting probability therefore transforms from a linear profile at $\gamma \tau = 0$ to an essentially flat profile for $\gamma \tau \gg 1$. Furthermore, we have shown in Fig.~\ref{fig-classical} that the splitting probability varies smoothly with the sampling time and exhibits no phase-transition-like behavior as observed for quantum case.
}

\end{widetext}

\end{document}